\documentclass{tlp}
\pdfoutput=1

\usepackage{xspace}
\usepackage{latexsym}
\usepackage{amssymb}   %
\usepackage{amsmath}

\usepackage{ifthen}
\newboolean{commentsaon}         %
\setboolean{commentsaon}{true}
  \setboolean{commentsaon}{false}  %

\newboolean{commentson} %
\setboolean{commentson}{true}

\newcommand{\comment}[1]
{\ifthenelse{\boolean{commentson}\AND\boolean{commentsaon}}
   {{\par\noindent\mbox{}{\small\blue[ *** #1 ]\par}\noindent\par}}{}}

\newcommand{\commenta}[1]
{\ifthenelse{\boolean{commentsaon}}
   {{\par\noindent\mbox{}{\small\color[rgb]{0, .5, 0}[ *** #1 ]\par}\noindent\par}}{}}
\usepackage[ddmmyyyy]{datetime}

\renewcommand{\today}{18-04-2017}

\newcommand{\myhalfmagnification}{0.003}

\usepackage{color}
\newcommand\blue     {\color{blue}}

\usepackage{hyperref}
\usepackage[hyphenbreaks]{breakurl}
\usepackage{pgf}

\newtheorem{theorem}{Theorem}
\newtheorem{corollary}[theorem]{Corollary}
\newtheorem{lemma}[theorem]{Lemma}
\newtheorem{proposition}[theorem]{Proposition}

\newtheorem{definition}[theorem]{Definition}

\newcommand*{\seq}[2][n]  {{#2_{1}, \allowbreak \ldots, \allowbreak #2_{#1}}}

\newcommand*{\mydash}{{\mbox{\tt-}}}
\newcommand*{\HU}{{\ensuremath{\cal{H U}}}\xspace}
\newcommand*{\HB}{{\ensuremath{\cal{H B}}}\xspace}

\newcommand*{\M}{{\ensuremath{\cal M}}\xspace}

\usepackage[mathscr]{euscript} 
\renewcommand*{\S}{{\ensuremath{\mathscr S}}\xspace}

\newcommand*{\NN}{{\ensuremath{\mathbb{N}}}\xspace}

\title[Logic + control]
{Logic + control:  \\ 
  On program construction and  verification}

\author[W. Drabent]
    {
    W{\l}odzimierz Drabent%
        \\        
         Institute of Computer Science,
         Polish Academy of Sciences
	\\
         IDA, Link\"oping University, Sweden \\
         {\tt drabent\,{\it at}\/\,ipipan\,{\it dot}\/\,waw\,{\it dot}\/\,pl}
}

\submitted{\dots }
\revised{\today }
\accepted{\dots }

\begin{document}
\maketitle

\begin{abstract}
This paper presents an example of formal reasoning about the semantics of a
Prolog program of practical importance (the SAT solver of Howe and King).
The program is treated as a definite clause logic program with added control.
The logic program is constructed by means of stepwise refinement, hand in
hand with its correctness and completeness proofs.
The proofs are declarative -- they do not refer to any
operational semantics. 
Each step of the logic program construction follows a systematic approach to
constructing programs which are provably correct and complete.
We also prove that correctness and completeness of the logic
program is preserved in the final Prolog program.
Additionally, we prove termination, occur-check freedom and non-floundering.

Our example shows how dealing with ``logic'' and with ``control'' 
can be separated. 
Most of the proofs
can be done at the ``logic'' level, abstracting from any operational semantics.

The example employs approximate specifications; they are crucial in
simplifying reasoning about logic programs.
It also shows that the paradigm of semantics-preserving program
transformations may be not sufficient.  
We suggest considering transformations which preserve correctness and
completeness with respect to an approximate specification.

\end{abstract}

\section{Introduction}

The purpose of this paper is to show that 1.~the correctness-related
issues of Prolog programs can, in practice, be dealt with mathematical
precision, 
and 2.~most of the reasoning can be declarative
(i.e.\ not referring to any operational semantics,
in other words depending only on the logical reading of programs%
).
We present a construction of a useful Prolog program.
We view it as a logic program with added control
\cite{DBLP:journals/cacm/Kowalski79}. 
The construction of the logic program is guided by
(and done together with)
a proof that the program conforms to its specification.
The Prolog program is obtained from the logic program by adding control.
We prove that adding control preserves the conformity with the specification.
The proofs follow the approach presented and discussed in
\cite{drabent.tocl16}. 
   We believe that the employed proof methods are not difficult and can be used
   in actual practical programming.

The notion of partial correctness of programs
(in imperative and functional programming) 
divides in logic programming into correctness and completeness.
Correctness means that all answers of the program are compatible with the specification,
completeness -- that the program  produces all the answers required by the
specification. 
A specification may be {\em approximate}\/:
 for such a specification
some answers are allowed, but not required to be computed.
The program construction presented in this paper illustrates 
the usefulness of approximate specifications.

For proving correctness we use the method of \cite {Clark79}.
The method should be well known, but is often neglected.
For proving completeness a method of \cite{drabent.tocl16}
is used.  It introduces a notion of {\em semi-completeness}\/;
semi-completeness and termination imply completeness.
We also employ an approach from \cite{drabent.tocl16} 
for proving that completeness is preserved under pruning of SLD-trees.
We use the sufficient condition for termination
from \cite{DBLP:journals/jlp/Bezem93},
and introduce a sufficient condition for non-floundering.

 We are interested in treating logic programming as a declarative paradigm,
 and in reasoning about programs declaratively, i.e.\ 
 independently from their operational semantics.
 Correctness, semi-completeness and completeness are declarative properties
 of programs.
To prove them this paper uses purely declarative methods.
 The employed sufficient condition for completeness of pruned SLD-trees
 abstracts, to a substantial extent, from details of the operational semantics.

The program dealt with in this paper is the SAT solver of 
Howe and King \citeyear{howe.king.tcs-shorter}.
It is an elegant and
concise Prolog program of 22 lines.  Formally it is not a logic program, as it includes
{\tt non{}var/1} and the if-then-else construct of Prolog;
 it was constructed as an
implementation of an algorithm, using logical variables and
coroutining.  The algorithm is
DPLL
\cite{DBLP:journals/cacm/DavisLL62},
 with watched literals and unit propagation 
(see \cite{handbook-SATsolvers,howe.king.tcs-shorter} and references therein). 
Here we look at the program from
a declarative point of view.
We show how it
can be obtained by adding control to a definite clause logic program.

We first present a simple logic program of four clauses, and then
modify it (in two steps) in order to obtain a logic program
on which the intended control can be imposed.
The construction of each program begins with a specification,
describing the relations to be defined by the program.
The construction is guided by a proof of the program's
correctness and semi-completeness,
and is performed hand in hand with the proof.
We formulate a general approach to such program construction.
The control imposed on the last logic program involves modifying the
selection rule (by means of delay declarations)
and pruning some redundant fragments of the search space.
Such control preserves correctness, and we
prove that completeness is also preserved
(which in general may be violated by pruning, or by floundering).

It is important
that logic and control are separated in the construction.
Most of the work has been done at the level of logic programs.
Their correctness and completeness could be treated formally, independently
from the operational semantics.
All the considerations related to the operational semantics,
program behaviour and efficiency
 are independent from those related to the declarative
semantics, correctness and completeness.

\paragraph{Related work.}
For an overview of work on proving correctness and completeness of logic
programs see \cite{drabent.tocl16,DBLP:journals/tplp/DrabentM05shorter}.  In
particular, little work has been done on program completeness.
The author is not aware of published correctness (or completeness)
proofs that are declarative, and deal with practical logic programs
(not smaller than the program dealt with here).%
\footnote{%
\label{footnote:nondeclarative}
  The proof methods employed in
  \cite{Apt-Prolog,DBLP:journals/jlp/PedreschiR99} are not declarative. 
  In particular, they depend on the order of atoms in clause bodies.
  They prove certain properties of LD-derivations, from which 
  program correctness, in the sense considered here, follows.
  (In the case of the latter paper, also program completeness
  can be concluded.)

  Similarly, the approach of \citeN{Deville} is not declarative.
  Also,
  it deals with constructing programs from specifications;
  correctness and completeness follow from construction.  Thus
  reasoning about correctness or completeness of arbitrary programs is not
  dealt with.  

}
A possible exception is a correctness proof of a toy compiler 
of 7 clauses \cite{Deransart.Maluszynski93}.

A preliminary version of this article was \cite{drabent12.iclp};
that paper presented basically the same program construction, however 
some proofs presented here were missing.  In particular preserving
completeness under pruning was dealt with only informally.
The methods for dealing with completeness and correctness are presented and
discussed in \cite{drabent.tocl16}; 
here we augment that paper by a more substantial example.

Our approach differs from semantics-preserving program transformations 
 \cite[and the references therein]{PettorossiPS10shorter}.
The latter is focused on formal steps for transforming programs into more
efficient ones with the same semantics (of their main predicates).
The initial program is treated as a specification.
In our example, the role of intermediate programs is to illustrate the
informal development process, and support explaining the design decisions.
Correctness and completeness are considered for each program separately.
An interesting feature of our example is that the consecutive programs
are not equivalent -- their main predicates define different
relations, satisfying however the same approximate specification.
We argue
(Section \ref{sec:transformations})
that the paradigm of semantics-preserving transformations is inapplicable
to this case.
This suggests that it should be useful to generalize the paradigm 
to transformations which preserve
correctness and completeness w.r.t.\ an approximate specification.

\paragraph{Preliminaries.}
This paper considers definite clause logic programs
(then implemented as Prolog programs).
We use the standard notation and definitions \cite{Apt-Prolog}.
As we deal with clauses as data, and clauses of programs, the latter will be
called {\em rules} to avoid confusion.
Given a predicate symbol $p$, by a {\em $p$-atom} (or {\em atom for} $p$)
 we mean an atom whose
predicate symbol is $p$, and by a {\em rule for} $p$ -- a rule whose head
is a $p$-atom.
The set of the rules for $p$ in the program under consideration is called
{\em procedure} $p$.

We assume a fixed alphabet of function and predicate symbols.
The alphabet may contain symbols not occurring in the considered program.
The Herbrand universe (the set of ground terms) will be denoted by \HU, the
Herbrand base (the set of ground atoms) by \HB. 
For an expression (a program) $E$
by $ground(E)$ we mean the set of ground instances
of $E$ (ground instances of the rules of $E$).
$\M_P$ denotes the least Herbrand model of a program $P$.
An expression is {\em linear} if every variable occurs in it at most once.

  By a computed (respectively correct) answer for a program $P$ and a
  query $Q$ we mean an instance $Q\theta$ of $Q$ where $\theta$ is a computed
  (correct) answer substitution \cite{Apt-Prolog} for $Q$ and $P$.
(So we use ``computed/correct answer'' instead of
 ``computed/correct instance of a query'' of \cite{Apt-Prolog}.)
  We often say just
   {\em answer}
 as each computed answer is a correct one, 
  and each correct answer (for $Q$) is a computed answer
  (for $Q$ or for some instance of $Q$).
  Thus, by soundness and completeness of SLD-resolution,
  $Q\theta$ is an answer for $P$ iff $P\models Q\theta$.
We also deal with a generalization of standard 
SLD-resolution, allowing the selection rule to be a partial function
(a {\em partial selection rule}).
Thus in some queries no atom may be selected, 
such a query is called {\em floundered}.
Obviously, such a generalization preserves correctness, but not completeness of
SLD-resolution. 
By the Prolog selection rule we mean selecting the first atom in each query.
(So most of Prolog implementations also implement other selection rules; this
is called delays or coroutining.)
LD-resolution means SLD-resolution under the Prolog selection rule.

By ``declarative'' (property, reasoning, \ldots) 
we mean referring only to the logical reading of programs, 
thus abstracting from any operational semantics.
So $Q$ being an answer for $P$ is a declarative property.
In particular,
properties depending on the order of atoms in rules will not be considered
declarative, as the logical reading does not distinguish equivalent formulae,
like 
$\alpha\land\beta$ and $\beta\land\alpha$.

  Names of variables begin with an upper-case letter.
  We use the list notation of Prolog.  So 
  $[\seq t]$  ($n\geq0$) stands for the list of elements $\seq t$.
  Only a term of this form is considered a list.
 (Thus terms like $[a,a|X]$, or  $[a,a|a]$ where $a$ is a constant distinct
  from $[\,]$, are not lists).
The set of natural numbers will be denoted by \NN.

\paragraph{Outline of the paper.}
Sections \ref{sec:corr-compl} and \ref{sec:compl-pruning}
present theoretical results showing how to, respectively, prove correctness
and completeness of programs, and prove that completeness is preserved under
pruning of SLD-trees.
Sections \ref{sec:logicProgram1} and
\ref{sec:program}
develop a SAT-solver as a logic program, hand in hand with its correctness
and completeness proof.
Additionally, the latter section proves non-floundering and occur-check
freedom, and discusses semantics-preserving transformations in the context
of our example.
Section \ref{sec:prologprogram}
converts the logic program into a Prolog program with the intended control
imposed.  It also shows that the Prolog program preserves the correctness and
completeness of the logic program.
The last section contains a discussion.

\section{Correctness and completeness of programs}
\label{sec:corr-compl}
This section introduces the notions of specification, correctness and
completeness.  Then it presents a way of proving that definite logic programs
are correct and complete.  The approach is declarative.  
It does not depend on any operational semantics, and
programs are viewed as sets of logic formulae.
In particular, the reasoning is independent from the order of atoms in rules.
We conclude with presenting a way of constructing programs which are provably
correct and semi-complete.
The presentation here is rather brief,
for a more comprehensive treatment see  \cite{drabent.tocl16}.
This section lacks examples,
as the concepts introduced here are employed later on in the paper.
In particular, Section \ref{sec:logicProgram1} provides examples 
for most of the definitions and results presented here.

\subsection{The notions}
\label{subsec:notions}

\paragraph{Specifications.}
From a declarative point of view, logic programs compute relations. 
A specification should describe these relations.
It is convenient to assume that the relations are over the Herbrand universe. 
A handy way for describing such relations is a Herbrand interpretation;
it describes, as needed, a relation for each predicate symbol of the program.
\begin{definition}
  \label{def:specification}
  By a {\bf specification} we mean a Herbrand interpretation, i.e.\ a
  subset of \HB.

  The relation described by a specification $S$ for a predicate $p$
  of arity $n$ is
  {$\{\, (t_1,\ldots,t_n)  \mid p(t_1,\ldots,t_n)\in S \,\}$}.
  The atoms from a specification $S$ will be called the 
  {\bf specified} atoms (by $S$).
  \sloppy
\end{definition}
Obviously,
 the relations actually defined by a program $P$ are described by its
least {Herbrand} model $\M_P$; the relation for a predicate $p$ is
{$\{\, (t_1,\ldots,t_n)  \mid p(t_1,\ldots,t_n)\in \M_P \,\}$.}

\paragraph{Correctness and completeness.}
In imperative and functional programming, (partial) correctness usually means
that the program results are as specified (provided the program terminates).
  In logic programming, due to its non-deterministic nature,
we actually have two issues: {\em correctness} (all the results are
compatible with the specification) and {\em completeness} (all the results
required by the specification are produced). 
In other words, correctness means that the relations defined by the program are
subsets of the specified ones, and completeness means inclusion in the
opposite  direction. 
Formally:
\begin{definition}
\label{def:corr:compl}
Let $P$ be a program and $S\subseteq\HB$ a specification.
$P$ is {\bf correct} w.r.t.\ $S$ when $\M_P\subseteq S$;
it is {\bf complete} w.r.t.\ $S$ when $\M_P\supseteq S$.
\end{definition}
We will sometimes skip the specification when it is clear from the context.
For a program $P$ correct w.r.t.\ $S$,
if a query $Q$ is an answer of $P$ then $S\models Q$.
(Remember that $Q$ is an answer of $P$ iff $P\models Q$.)
When $P$ is complete w.r.t.\ $S$ and $Q$ is ground,
then $S\models Q$ implies that $Q$ is an answer of $P$. 
More generally, this implication holds for arbitrary queries such that
$\M_P\models Q$ implies $P\models Q$
(for sufficient conditions,
see \cite{DBLP:books/mk/minker88/Maher88,Apt-Prolog,drabent.tocl16,drabent.Herbrand.2016}).
In particular, the implication holds when the alphabet of function symbols is infinite.

\pagebreak[3]
It is sometimes useful to consider local versions of these notions:

\pagebreak[3]
\begin{definition}
\nopagebreak
\label{def:corr:compl:local}
\nopagebreak
A {\bf predicate} $p$ in $P$ is
 {\bf correct} w.r.t.\ $S$ when each $p$-atom of $\M_P$ is in $S$, and 
 {\bf complete} w.r.t.\ $S$ when each $p$-atom of $S$ is in $\M_P$.

An {\bf answer} $Q$ is  {\bf correct} w.r.t.\ $S$ when $S\models Q$.

$P$ is {\bf complete for a query} $Q$ w.r.t.\  $S$
when
$S\models Q\theta$ implies that $Q\theta$ is an answer for $P$,
for any ground instance $Q\theta$ of $Q$.
\end{definition}
Informally,  $P$ is complete for $Q$
when 
all the answers for $Q$ required by the specification $S$ are answers of $P$.
Note that a program is complete w.r.t.\ $S$
 iff it is complete w.r.t.\ $S$ for any query
 iff it is complete w.r.t.\ $S$ for any query $A\in S$.

\ifthenelse{\boolean{commentsaon}}{\pagebreak[3]}{}

\paragraph{Approximate specifications.}
It happens quite often in practice that a programmer
does not know exactly the relations defined by a program
and, moreover, such knowledge is unnecessary.
For example, it is irrelevant whether 
$ {\it append}([a], 1, [a|1])$ is an answer of a list appending program or not.
In such cases, it is
 sufficient to specify the program's semantics approximately.
More formally, to provide distinct specifications, say
$S_{\it c o m p l}$ and $S_{corr}$, for completeness and correctness
of the considered program $P$.
The intention is that  $S_{\it c o m p l}\subseteq \M_P\subseteq S_{corr}$.
So the specification for completeness says what the program has to compute,
and the specification for correctness -- what it may compute;
in other words, the program should not produce any answers
incorrect w.r.t.\ the specification for correctness. 
It is irrelevant whether atoms from 
$S_{corr}\setminus S_{\it c o m p l}$ are, or are not, answers of the program.
For example the standard \mbox{APPEND} program does not define the list
concatenation relation, but its superset
(and $ {\it append}([a], 1, [a|1])$ is an answer of the program,
but $ {\it append}([a], a, [a|1])$ is not).
In this case, $S_{\it c o m p l}$ describes the list concatenation relation;
see \cite{drabent.tocl16}
for  $S_{corr}$ and further discussion.

\begin{definition}
An {\bf approximate specification} is 
a pair $S_{\it c o m p l}, S_{corr}$ of specifications, where
$S_{\it c o m p l}\subseteq S_{corr}\subseteq \HB$.

A program $P$ is {\bf fully correct} w.r.t.\  $S_{\it c o m p l}, S_{corr}$
when  $S_{\it c o m p l}\subseteq \M_P\subseteq S_{corr}$.
A predicate $p$ in $P$ is fully correct w.r.t.\ $S_{\it c o m p l}, S_{corr}$
when each $p$-atom of $S_{\it c o m p l}$ is in $\M_P$,
and each $p$-atom of $\M_P$ is in $S_{corr}$.
\end{definition}
Correctness (respectively completeness) w.r.t.\ 
$S_{\it c o m p l}, S_{corr}$ will mean
correctness w.r.t.\ $S_{corr}$ (completeness w.r.t.\ $S_{\it c o m p l}$). 
By abuse of terminology,
a single specification may be called
approximate when the intention is that the specification is
distinct from~$\M_P$.

In many cases, the atoms from $S_{corr}\setminus S_{\it c o m p l}$ 
may be (formally or informally) considered as ill-typed.
For instance, in ${\it append}([a], 1, [a|1])$
terms $1$ and $[a|1]$ are not lists. 
In the context of typed logic programming \cite{types.lp.92-short},
the need for approximate specifications may disappear.
For instance, in an appropriate
typed logic the standard APPEND program may define the list concatenation exactly.
However in practice we usually deal with untyped logic programming.
In particular, the programming language Prolog is untyped.
See \cite{drabent.tocl16} for further discussion, including the need
for approximate specifications in typed logic programming. 

The main example of this paper employs approximate specifications.
In particular,
  various versions of the constructed program define different relations
  for the same predicate symbols, but the programs 
  are correct and complete w.r.t.\ the same approximate specification.
(More precisely, their common predicates are
fully correct w.r.t.\ the same approximate specification.)
The example shows that employing approximate specifications
results in simplifying the construction of specifications and proofs.
It is useful \cite{drabent.tocl16} in declarative diagnosis
(also known as algorithmic debugging \cite{Shapiro.book}).
For further discussion and examples see 
\cite{drabent.tocl16,DBLP:journals/tplp/DrabentM05shorter}.  

\subsection{Reasoning about correctness}
The following sufficient condition for program correctness
will be used.
\begin{theorem}
\label{th:correctness}
Let $P$ be a program and $S$ be a specification.
If  $S\models P$  then  $P$ is correct w.r.t.\ $S$.
\end{theorem}
\pagebreak[3]
In other words, the sufficient condition for correctness of $P$ is that
     for each ground instance 
    $
    H\gets \seq B
    $
    of a rule of $P$,
     if $\seq B\in S$ then $H\in S$.

\citeN{DBLP:journals/tcs/Deransart93} attributes this result to \cite{Clark79}.
It  should be well known, but is often unacknowledged.
Often more complicated correctness proving methods, based on the operational
semantics, are proposed, e.g.\ in  \cite{Apt-Prolog}.
See \cite{DBLP:journals/tplp/DrabentM05shorter,drabent.tocl16}
for further comparison, examples and discussion.

\subsection{Reasoning about completeness}
\label{subsec:completeness}
Little work has been devoted to reasoning about completeness of programs.
See \cite{drabent.tocl16} for an overview.  We summarize the approach
of \cite{drabent.tocl16}, stemming from that of
\cite{DBLP:journals/tplp/DrabentM05shorter}.
It is based on an auxiliary notion of semi-completeness.

\pagebreak[3]
\begin{definition}
\nopagebreak
A program $P$ is {\bf semi-complete} 
w.r.t.\ a specification $S$ if
$P$ is complete w.r.t.\ $S$ for any query $Q$
such that there exists a finite SLD-tree for $P$ and $Q$.
\end{definition}

Less formally, the existence of a finite SLD-tree means
that $P$ with $Q$ terminates under some selection rule
($\exists$-terminates in the terms of \cite{PedreschiRS02.TPLP.terminating}).
For a semi-complete program $P$, if a computation for a query $Q$
terminates then all the answers for $Q$ required by the specification have
been obtained.
So establishing completeness is divided into showing completeness and
termination. 
Obviously, a complete program is semi-complete.
We immediately obtain:
\begin{proposition}[completeness]
\label{prop:completeness:trivial}
Assume that a program $P$ is semi-complete w.r.t.\ a specification $S$.
Then $P$ is complete w.r.t.\  $S$ if
  \begin{enumerate}
  \item each atom $A\in S$ is a root of a finite SLD-tree for $P$, or
  \item each atom $A\in S$ is an instance of an atom $Q$ which is a root of a
    finite SLD-tree for $P$.
  \end{enumerate}
\end{proposition}
\pagebreak[3]

Our sufficient condition for semi-completeness employs the
following notion, stemming from \cite{Shapiro.book}.
\begin{definition}
  A ground atom $H$ is
  {\bf covered by a rule} $C$ w.r.t.\ a specification $S$
  if $H$ is the head of a ground instance  
  $
  H\gets \seq B
  $
  ($n\geq0$) of $C$, such that all the atoms $\seq B$ are in $S$.

  A ground atom $H$ is {\bf covered} {\bf by a program} $P$ w.r.t.\ $S$
  if it is covered w.r.t.\ $S$ by some rule $C\in P$.
\end{definition}

Informally, $H$ covered by $C$ w.r.t.\ $S$ means that $C$ can produce $H$ out
of the atoms in $S$.
The following sufficient condition provides a method of proving
semi-completeness. 

\begin{theorem}[semi-completeness {\rm\cite{drabent.tocl16}}]
\label{th:semi-complete}
If all the atoms from a specification $S$ are covered w.r.t.~$S$
 by a program $P$ 
then $P$ is semi-complete w.r.t.~$S$.
\end{theorem}

We will say that a procedure $p$ of $P$
{\em satisfies the sufficient condition for semi-completeness} w.r.t.\ $S$
if each $p$-atom of $S$ is covered by $P$ w.r.t.\ $S$.

To prove program completeness, the theorem has to be augmented by a way of
proving termination.  Fortunately, for the purposes of this paper 
a simple method of \cite{DBLP:journals/jlp/Bezem93} is sufficient.

\begin{definition}
A {\bf level mapping} is a
function $|\ |\colon \HB\to \NN$ assigning natural numbers to ground atoms.

A program $P$ is  {\bf recurrent} {w.r.t.\ a level mapping}~$|\ |$
\cite{DBLP:journals/jlp/Bezem93,Apt-Prolog} if, in
every ground instance  $H\gets\seq B\in ground(P)$ of its rule ($n\geq0$),
$|H|>|B_i|$ for all $i=1,\ldots,n$.

A program is {recurrent}
if it is recurrent w.r.t.\ some level mapping.   
A rule $C$ is {recurrent} (w.r.t.\ $|\ |$) if program $\{C\}$
is {recurrent} (w.r.t.\ $|\ |$).

A query $Q$ is {\bf bounded} w.r.t.\ a level mapping  $|\ |$ if, 
for some $k\in\NN$, $|A|<k$ for each ground instance $A$ of an atom of $Q$.
\end{definition}
 \vspace{-4pt}
\begin{theorem}
[termination {\rm\cite{DBLP:journals/jlp/Bezem93}}]
\label{th:termination}
Let $P$ be a program recurrent w.r.t.\ a level mapping $|\ |$,
and $Q$ be a query bounded w.r.t.\ $|\ |$.
Then all SLD-derivations for $P$ and $Q$ are finite.
\end{theorem}

From the theorem and Proposition \ref{prop:completeness:trivial} we
immediately obtain: 
\begin{corollary}
[completeness]
\label{cor:completeness:recurrent}
If a program $P$ is semi-complete w.r.t.\ a specification $S$ and
recurrent
then $P$ is complete w.r.t.~$S$.
\end{corollary}

Note that the sufficient conditions for correctness (Theorem \ref{th:correctness})
and semi-completeness (Theorem \ref{th:semi-complete}) are declarative.  So is
the sufficient condition for completeness of Corollary
\ref{cor:completeness:recurrent}.
Thus the correctness and completeness proofs presented later on in this paper
are declarative. 
In general, the presented approach for proving completeness is not declarative,
as termination (finiteness of SLD-trees) depends on the selection rule.
For instance, the well-known method of \citeN{AP93} of proving termination,
depends on the order of atoms in program rules.
{\sloppy\par}

We mention another declarative way of showing program
completeness \cite{Deransart.Maluszynski93}
(see also \cite{drabent.tocl16}),
applicable also to non-terminating programs.
In that approach a level mapping is employed, however it may be defined only
on atoms 
from the specification $S$.
\begin{theorem}
[completeness]
\label{th:completenessDeransart}
Let $P$ be a program, $S$ a specification, and $|\ |\colon S\to \NN$.
If each atom $A\in S$ is covered w.r.t.\ $S$ by some ground
instance $A\gets\seq B \in ground(P)$ such that 
$|A|> |B_i|$ for $i=1,\ldots,n$
then $P$ is complete w.r.t.\ $S$.
  
\end{theorem}

Note the similarity of this condition to that of Theorem \ref{th:semi-complete}
together with  Corollary \ref{cor:completeness:recurrent}.
The difference is that here only a fragment of $ground(P)$ is required to be
recurrent.  The fragment consists of a rule instance covering $A$ for each
$A\in S$.\linebreak[3]
Due to this similarity, the amount of work to prove program completeness by 
Theorem \ref{th:completenessDeransart}
is often similar to that of proving it 
by showing semi-completeness and termination.
So in practice the latter is usually preferable, as the termination 
has to be shown anyway
(under the selection rule used by the implementation of the program).

\subsection{Program construction}
\label{sec:construction}
The presented sufficient conditions suggest a systematic (informal) method
of constructing programs which are provably correct and semi-complete.
A guiding principle is that the program should satisfy the sufficient
condition for semi-completeness.
The construction results in a program together with proofs of its
correctness and semi-completeness.

Assume that specifications $S_{\it c o m p l}$ and $S_{corr}$ are given 
for, respectively, completeness and correctness of a program $P$ to be built. 
For each predicate $p$ occurring in $S_{\it c o m p l}$ consider the set
$S_p = \{\, p(\bar t)   \mid p(\bar t) \in S_{\it c o m p l} \,\}$
of the specified $p$-atoms from the specification.
To construct a procedure $p$ of $P$, provide rules such that
\begin{enumerate}
\item 
\label{requirement1}
  each atom $A\in S_p$ is covered w.r.t.\ $S_{\it c o m p l}$ by some rule, and
\item
  each rule satisfies the sufficient condition for correctness
  w.r.t.\ $S_{corr}$ of Theorem~\ref{th:correctness}.
\end{enumerate}
(In other words, the first requirement states that the constructed procedure
{satisfies the sufficient condition for semi-completeness}.)
The constructed program $P$ is the union of the procedures for all $p$ from
$S_{\it c o m p l}$.  
It satisfies the conditions of Theorems \ref{th:correctness},
\ref{th:semi-complete}.  Thus $P$ is correct w.r.t.\ $S_{\it corr}$ and
semi-complete w.r.t.\ $S_{\it c o m p l}$.

In practice, semi-completeness is not sufficient.  The actual task is 
to obtain a program which is complete;  also in most cases the program should
terminate for the intended class of initial queries.   
So the constructed program rules
should additionally satisfy some sufficient condition
for completeness (like that of Theorem \ref{th:completenessDeransart}) or
for termination
(like the program being recurrent).
In this way the method of the previous paragraph may be augmented to ensure
not only correctness and semi-completeness, but also completeness.

The approach presented here will be used throughout the program constructions
presented in this paper.

\section{SAT solver -- first logic program}
\label{sec:logicProgram1}

We are ready to begin the main subject of this paper -- a construction of a
 program implementing a SAT solver.
The construction is divided in several steps, three definite clause logic
programs and a final Prolog program are constructed.
An interesting feature is that the construction is not a case of
semantics-preserving program transformation. 
The programs define different relations (for
the common predicates); however the common predicates are correct and
complete w.r.t.\ the same approximate specification.

This section explains the data structures used by the programs, 
provides a specification, 
and presents a construction of the first program, hand in hand with 
a correctness and semi-completeness proof.

\paragraph{Representation of propositional formulae.}

We first describe the form of data used by our programs, namely
the encoding of propositional formulae in CNF as terms,
proposed by \cite{howe.king.tcs-shorter}.

Propositional variables are represented as logical variables;
truth values -- as constants {\tt true}, {\tt false}.
A literal of a clause is represented as a pair of a truth value and a variable;
a positive literal, say $x$, as {\tt true-X}
and a negative one, say $\neg x$, as  {\tt false-X}.
A clause is represented as a list of (representations of) literals,
and a conjunction of clauses as a list of their representations.
For instance a formula
$(x\lor\neg y\lor z)\land(\neg x\lor v)$ is represented as 
{\tt[[true-X,false-Y,true-Z],[false-X,true-V]]}.

An assignment of truth values to variables can be represented as a substitution.
Thus a clause (represented by a term) $s$ is true under an assignment (represented
by) $\theta$ iff the list $s\theta$ 
has an element of the form  $t\mydash t$, i.e.\ 
{\tt false-false} or {\tt true-true}.  A formula in CNF is satisfiable
iff
its representation has an instance whose
elements (being lists) contain a  $t\mydash t$ each.
We will often say ``formula~$u$\/'' for a formula in CNF represented as a
term $u$, similarly for clauses etc.

\paragraph{Specification.}
Now let us describe the sets to be defined by the predicates of our first 
SAT-solving program.  
\begin{equation}
\label{L1L2}
  \begin{array}{l}
 L_1 = \{\, [\seq[i]t|s]\in\HU \mid  i>0,\ 
            \mbox{$t_i = t\mydash t$ for some $t\in\HU$} \,\},
    \\
    L_2 = \{\, [\seq s]\in\HU \mid n\geq0,\ \seq s\in L_1 \,\}.
  \end{array}
\end{equation}
Note that $t_1,\ldots,t_{i-1},s$ are arbitrary ground terms; the reason for
such generality will be explained further on.
The following holds for $L_1$ and $L_2$.
\begin{equation}
\label{property.spec}
  \begin{array}{l}
    \mbox{A clause $s$ is true under an assignment $\theta$ iff the list
      $s\theta$ is in $L_1$. Hence}
    \\
    \mbox{a CNF formula $u$ is true under $\theta$ iff $u\theta$ is
      in $L_2$,} 
    \\
    \mbox{a CNF formula $u$ is satisfiable iff $u$ has an instance in $L_2$.} 
  \end{array}
\end{equation}
Let us require that predicates {\it sat\_cl}, ${\it sat\_c n f}$ define
in our program the sets $L_1,L_2$.
So the specification is 
\[
S_1 = \{\, {\it sat\_cl}(s) \mid s\in L_1\,\}
 \cup \{\, {\it sat\_c n f}(u) \mid u\in L_2\,\}.
\]
From  (\ref{property.spec}) it follows that each program
which is correct and complete w.r.t.\ $S_1$ works as a SAT solver:
for any CNF formula $u$ and any (substitution representing an) assignment
$\theta$,
\begin{equation}
\label{property.program}
{%
\hspace*{-2em}
    \begin{tabular}{@{}l@{}}
      $u$ is true under $\theta$ iff
      $\theta$ is an answer substitution for ${\it sat\_c n f}(u)$,     
      \\
      $u$ is satisfiable iff query ${\it sat\_c n f}(u)$ succeeds.
    \end{tabular}%
\hspace*{-2em}
}
\end{equation}

We are ready with the specification for our first program.
However, it is not necessary that a SAT-solving program defines the set $L_2$.
Another choice may be
\begin{equation}
\label{L10}
\begin{array}{l}
    L_1^0 = 
     \left
    \{\, 
    \rule{0pt}{2.ex}
    [t_1\mydash t'_1,\ldots, t_n\mydash t'_n]  \in\HU
     \ \left| \ 
    \rule{0ex}{2ex}
    n>0,
    \
    t_i=t'_i \mbox{ for some } i\in \{1,\ldots,n\} %
     \right.\right
    \},
  \\
        L_2^0 = \{\, [\seq s]\in\HU \mid n\geq0,\ \seq s\in L_1^0 \,\},
\end{array}
\end{equation}
as (\ref{property.spec}) holds for $L_1^0, L_2^0$.
(Note that  $L_1^0\subseteq L_1$ and $L_2^0\subseteq L_2$;
for simplicity, we allow arbitrary terms $t_i,t'_i$, not only 
{\tt false}, {\tt true}.)
This leads to a specification
\[
S_1^0 = \{\, {\it sat\_cl}(s) \mid s\in L_1^0\,\}
 \cup \{\, {\it sat\_c n f}(u) \mid u\in L_2^0\,\}.
\]  
Moreover, any sets 
$L_1'$, $L_2'$ such that $L_1^0\subseteq L_1'\subseteq L_1$
and $L_2^0\subseteq L_2'\subseteq L_2$ will do,
as such $L_1'$, $L_2'$ satisfy (\ref{property.spec}).%
\footnote{Because
   if a clause $s$ has an instance $s\theta\in L_1$ then $s\theta\in L_1^0$,
   and
   if a CNF formula $u$ has an instance $u\theta\in L_2$ then $y\theta\in L_2^0$.
} %
Hence any program complete w.r.t.\ $S_1^0$ and correct w.r.t.\ $S_1$ has
property (\ref{property.program}), and thus can be used as a SAT solver.

We have chosen $L_2$  as the set defined by our first program
(so $S_1$ is its specification both for correctness and for completeness).
  This leads to a simpler program
-- a check is avoided that certain terms are lists, and that their elements
are of the form $t\mydash t'$.
However our further programs will define a set $L_2'$ as above,
employing $S_1^0, S_1$ as an approximate specification
(for predicates  ${\it sat\_cl}, {\it sat\_c n f}$).%
\footnote{%
    Our choice of the approximate specification is rather arbitrary.
    $S_1^0$ could be shrunk by additionally requiring
    in the definition of $L_1^0$
         that $t_j,t'_j\in\{ {\tt true},{\tt false} \}$
            for $j=1,\ldots, n$.
        $S_1$ can be extended by setting 
        $L_1= \{\, t\in\HU \mid 
        \mbox{if } t \mbox{ is of the form } 
        [t_1\mydash u_1,\ldots,t_n\mydash u_n] \mbox{ then }
        t_i\mathop=u_i \mbox{ for some } i \,\}
        $ 
        \cite{drabent12.iclp}.
  However, the chosen $S_1^0, S_1$ seem more convenient.

  Let us also note that in a suitable typed logic 
  the need for an approximate specification
  would disappear for this particular problem.  Roughly speaking,
  if the argument of ${\it sat\_c n f}$ is restricted to the type of
  (representations of) CNF formulae then 
  the set to be defined by ${\it sat\_c n f}$ is unique.
  The set is specified by the shrunk $S_1^0$, described above
  in this footnote.
} %

\paragraph{The first program.}
Now we construct a program $P_1$, hand in hand with its correctness and
semi-completeness proofs.
We follow the program construction approach described in Section
\ref{sec:construction}.
So we construct program rules such that
each atom from $S_1$ is covered w.r.t.\ $S_1$ by some of the rules.
Additionally, care is taken that the rules are recurrent.
(This is not made explicit in our text.)
For each rule the sufficient condition for correctness of 
Theorem \ref{th:correctness} will be checked.
At the end we prove that the constructed program is recurrent, 
hence complete and terminating.

By (\ref{L1L2}),
a ground {\it sat\_cl}-atom is in $S_1$ iff it is of the form
$A_1 = {\it sat\_cl}([t\mydash t|s'])$ or
$A_2 = {\it sat\_cl}([t|s])$ where $s\in L_1$.
For the first case, we provide a rule%
\begin{equation}
  sat\_cl([Pol\mydash {\it Pol} | {\it Pairs}]). 
  \label{satcl1}
\end{equation}
Obviously the rule covers each such  $A_1$, and the sufficient condition for
correctness w.r.t.\ $S_1$ (Theorem \ref{th:correctness}) is satisfied
(as each ground instance $sat\_cl([p\mydash p|s])$ of
(\ref{satcl1}) is in $S_1$).
For the second case, a rule
{\sloppy
    \begin{equation}
          sat\_cl([  H            | {\it Pairs}]) \gets 
            sat\_cl({\it Pairs}).
    \label{satcl2}
    \end{equation}
}%
covers each such $A_2$
(as $A_2$ is the head of a ground instance
$ {\it sat\_cl}([t | s]) \gets {\it sat\_cl}(s) $ of (\ref{satcl2}),
where  ${\it sat\_cl}(s)\in S_1$).
The sufficient condition for correctness holds for each ground instance of 
(\ref{satcl2}), as  ${\it sat\_cl}(s')\in S_1$ implies 
$s'\in L_1$, hence $[t'|s']\in L_1$, and ${\it sat\_cl}([t'|s'])\in S_1$.

A ground ${\it sat\_c n f}$-atom is in $S_1$ iff it is $B_1=[\,]$ or of the form
$B_2 = {\it sat\_c n f}([s|u])$ where $s\in L_1$ and $u\in L_2$.
The following rules cover  $B_1$, respectively $B_2$ w.r.t.\ $S_1$, 
(simple details are left to the reader).
\begin{eqnarray}
&&
{\it sat\_c n f}([\,]). \label{satcnf1}  \\
&&
{\it sat\_c n f}([Clause | Clauses]) \gets
    sat\_cl(Clause),
    \ {\it sat\_c n f}(Clauses). \label{satcnf2}  
\end{eqnarray}
If $sat\_cl(s),  {\it sat\_c n f}(u) \in S_1$ then $s\in L_1$, $u\in L_2$,
hence ${\it sat\_c n f}([s |u])\in S_1$.
So the sufficient condition of Theorem \ref{th:correctness}
holds for rule (\ref{satcnf2}).  It obviously holds for rule (\ref{satcnf1}).
So we constructed program $P_1$, consisting of rules
(\ref{satcl1}), (\ref{satcl2})
(\ref{satcnf1}), (\ref{satcnf2}), and proved that it is semi-complete and
correct w.r.t.\ $S_1$.

\paragraph{Termination and completeness of $P_1$.}

To show that the program is complete we show that it is recurrent
(cf.\ Corollary \ref{cor:completeness:recurrent}).
Let us define a level mapping  \mbox{$|\ |\colon \HB\to \NN$}
(and an auxiliary mapping  $|\ |\colon \HU\to \NN$ for terms):
\[
\begin{array}{l}
  |\, [h|t]\, | = |h|+|t|,    \\
  |f(\seq t)| = 1 \ \mbox{ where $n\geq0$ and $f$ is not }[\ | \ ],    \\
  |  {\it sat\_c n f}(t) | = | sat\_cl(t)|  = |t|,
    \\
\end{array}
\]
for any ground terms $h,t,t',\seq t$, and any function symbol $f$. 
Note that $|[\seq t]| = 1+ \Sigma_{i=1}^n |t_i|$, and that
$|t|>0$ for any term $t$.
It is easy to show that the program $P_1$ is recurrent under the level
mapping $|\ |$, i.e.\ 
for each ground instance $H\gets\ldots,B,\ldots$ of a rule of  $P_1$,
we have $|H|> |B|$.  For example, for a ground instance 
${\it sat\_c n f}([s|u])\gets {\it sat\_cl}(s),{\it sat\_c n f}(u)$
of (\ref{satcnf2}) we have
$|{\it sat\_c n f}([s|u])| = |s|+|u|$, which is greater both than
${\it sat\_cl}(s) =  |s|$ and ${\it sat\_c n f}(u)= |u| $.
(We leave further details to the reader.)
By Corollary \ref{cor:completeness:recurrent}, $P_1$ is complete w.r.t.\ $S_1$.

As a side effect, we obtain termination of $P_1$ for the intended queries
(as for any CNF formula $t$, $|t\theta|=|t|$ for any its instance $t\theta$;
thus query $Q={\it sat\_c n f}(t)$ is bounded  w.r.t.\ $|\,|$, 
and by Theorem \ref{th:termination} $P_1$~terminates for $Q$).

\paragraph{Summary.}
We chose, out of a few possibilities, the specification $S_1$ of a SAT solver
program, and then constructed such a program, namely $P_1$.
The construction was guided by the sufficient condition for semi-completeness
and performed hand in hand with a semi-completeness and correctness proof.

Note that the program is not correct w.r.t.\ $S_1^0$ (another 
specification considered here)
as, for instance, ${\it sat\_cl}([a,a\mydash a])\not\in S_1^0$ is an answer
of $P_1$. 
The reader is encouraged to construct a program correct and complete w.r.t.\ 
$S_1^0$,
in order to see that the program is more complicated (and most likely less
efficient), as it contains additional checks 
(that the argument of {\it sat\_cl} is a list and each its element is of the
form $t\mydash u$).

\section{Towards adding control}
\label{sec:program}

To be able to influence the control of program $P_1$ in the intended way,
in this section
we construct a more sophisticated logic program $P_3$, with a program 
$P_2$ as an intermediate stage.
(An impatient reader may find
$P_3$ on page \pageref{final.program}.)
The construction is guided by a formal specification, and done together with
a correctness and semi-completeness proof.%
\
It follows the approach of Section \ref{sec:construction}.
  We only partially discuss the reasons for particular design decisions in
  constructing the logic programs
  and in adding control, as the algorithmic and efficiency
  issues are outside of the scope of this work.
The constructed program is proved to be recurrent, from which its
completeness and termination follow.
The last two subsections concern occur-check freedom of $P_3$,
and inapplicability of correctness-preserving transformations to our example.

In this section not only the program, but also the specifications
are built incrementally
Our proofs will
remain valid under such changes of specifications by a rather obvious
property: 
\begin{lemma}
  Let $S'\subseteq S''\subseteq\HB$ be specifications and $P$ a program.

  If an atom $A$ is covered by $P$ w.r.t.\ $S'$ then it is covered by
  $P$ w.r.t.~$S''$. 

  If $P$ does not contain any predicate symbol occurring in
  $S''\setminus S'$, and
  the sufficient condition for correctness
  from Theorem \ref{th:correctness}
  holds for $P$ w.r.t.\ $S'$ 
   then 
  the condition holds for $P$ w.r.t.\ $S''$.
\end{lemma}

Program $P_1$ is correct and complete w.r.t.\ $S_1$.  However,
as explained in Section \ref{sec:logicProgram1}, 
   it is sufficient that predicate ${\it sat\_c n f}$ 
   is correct w.r.t.\ $S_1$ and complete w.r.t.\ $S_1^0$.
So now our program construction will be guided by an approximate
specification.   
Predicates  ${\it sat\_c n f}$, and ${\it sat\_cl}$ are going to be
correct w.r.t.\ $S_1$ and complete w.r.t.~$S_1^0$.

In what follows, SC1 stands for the sufficient condition
for correctness,
and SC2 stands for the fact that the considered procedure satisfies the
sufficient condition for semi-completeness
(cf.\ Theorem \ref{th:correctness} and the comment after
Theorem \ref{th:semi-complete}).
Let SC stand for SC1 with SC2.
We leave to the reader a simple check of SC2 for ${\it sat\_c n f}$
and $S_1^0$ (in Section \ref{sec:logicProgram1} this was done for  $S_1$).

\subsection{Preparing for adding control}
\label{sec:P2}

Program $P_1$ performs inefficient search by means of backtracking.
We are going to improve the search by changing the control.
The intended control cannot be applied directly to $P_1$;
so in this section we transform  $P_1$ to a program $P_2$.
The construction of $P_2$ has two aspects, formal and informal.
Formally, we extend the
specification in some steps (to describe new predicates), and apply the
augmented specifications to construct the rules of $P_2$
together with proofs of semi-completeness and correctness.
At an informal level, the construction of the specifications
(and to a certain extent of the program rules) is guided by the
intended control, to be eventually added to the program in Section
\ref{sec:prologprogram}.

Program $P_2$ includes the rules for ${\it sat\_c n f}$ from $P_1$,
i.e.\,(\ref{satcnf1}),\,(\ref{satcnf2}).
It contains a new definition of $sat\_cl$, and some new predicates.
The new predicates and $sat\_cl$ would define the same set $L_{sat\_cl}$
(or its subset  $L_{sat\_cl}\setminus \{\, [t]\mid t\in\HU\,\}$),
where $L_1^0 \subseteq L_{sat\_cl} \subseteq L_1$
(cf.\,(\ref{L1L2}),\,(\ref{L10})).
However they would represent elements of  $L_{sat\_cl}$ in a different way.
We are going to improve the search by delaying certain actions.
Invoking  $sat\_cl(s)$ is to be delayed when $s$
is a clause with the two first literals non-ground.
      (Note that a literal being non-ground means that it has no logical
      value assigned.)
In this way the technique of watched literals (cf.\ \cite{handbook-SATsolvers})
will be implemented. 
Unification of a non-ground pair $p\mydash v$ (by applying rule (\ref{satcl1})) 
is to be performed only if the argument of ${\it sat\_cl}$ is a unit clause,
that is $[p\mydash v]$.
  This contributes to implementing unit propagation
(cf.\ \cite{handbook-SATsolvers}).
Removing a literal from a clause (done in $P_1$ by rule (\ref{satcl2})) is to
be performed only when the literal is ground.
Also, if this literal is true then further search is unnecessary for this
clause.
This idea will be implemented 
 by separating two cases: the clause has one literal, or more.
For efficiency reasons we want to distinguish these two cases by means of
indexing the main symbol of the first argument of a predicate.
So the argument should be the tail of the list.  (The main symbol is $[\,]$
for a one element list, and $[\;|\;]$ for longer lists.)
We redefine  $sat\_cl$, 
introducing an auxiliary predicate $sat\_cl3$.
It defines the same set as $sat\_cl$, but a clause 
$[Pol\mydash {\it Var} | {\it Pairs}]$ is represented as three arguments 
 of  $sat\_cl3$.
More formally, its specifications for respectively correctness and
completeness are
\[
\begin{array}{l}
  S_{sat\_cl3} = \{\, sat\_cl3(s,v,p) \mid   [p\mydash v|s]\in L_1 \,\},
\\
  S_{sat\_cl3}^0 = \{\, sat\_cl3(s,v,p) \mid [p\mydash v|s]\in L_1^0 \,\}.
\end{array}
\]
A new procedure for  $sat\_cl$ is obvious:
\begin{equation}
\begin{array}{l}
sat\_cl([Pol\mydash {\it Var} | {\it Pairs}]) \gets sat\_cl3({\it Pairs}, {\it Var}, Pol). 
\end{array}
\label{satcl}
\end{equation}
SC are trivially satisfied
(w.r.t.\ specifications $S_{11}=S_1\cup S_{sat\_cl3}$,
and $S_{11}^0= S_1^0\cup S_{sat\_cl3}^0$,
we leave the simple details to the reader).%

Following \cite{howe.king.tcs-shorter} we will use in the program
an auxiliary predicate $=$ defining equality.  It will turn out necessary
when the final Prolog program is constructed.  Its specification 
(for correctness and completeness)
is obvious:
\[
S_= = \{\, t=t \mid t\in \HU\,\}.
\setlength\belowdisplayshortskip{3pt plus 1pt minus 7pt}
\]
As its definition, rule
\begin{equation}
\setlength\abovedisplayshortskip{3pt plus 1pt minus 7pt}
   {=}(X,X).
\label{ruleeq}  
\end{equation}
will do
(as it covers each atom from $S_=$, and satisfies the sufficient condition 
for correctness of Theorem \ref{th:correctness}.)

Procedure  $sat\_cl3$ has to cover each atom 
$A \in S_{sat\_cl3}^0$
i.e.\ each $A = sat\_cl3(s,v,p)$ such that
$[p\mydash v|s] = [t_1\mydash u_1,\ldots,t_n\mydash u_n]$ and $t_i=u_i$ for
some $i$.
Assume first $s=[\,]$.  Then $p=v$; this suggests a rule
\begin{equation}
    sat\_cl3([\,], {\it Var}, Pol) \gets {\it Var} = Pol. 
    \label{satcl31}
\end{equation}
Its ground instance 
\ $sat\_cl3([\,], p, p)  \gets p\mathop{=} p$ \ 
covers $A$ w.r.t.\ $S_{11}^0\cup S_=$.  Conversely, each instance of  (\ref{satcl31})
with the body atom in $S_{11}\cup S_=$ is of this form, its head is in $S_{11}\cup S_=$,
hence SC1 holds.

When the first argument of $sat\_cl3$ is not $[\,]$, then we want
to delay $sat\_cl3({\it Pairs}, {\it Var}, Pol)$ until ${\it Var}$ or the first
variable of ${\it Pairs}$ is bound.
In order to do this by means of {\tt block} declarations of SICStus
\cite{sicstus.Carlsson.tplp12},
we need to make the two variables to be separate arguments of a
predicate.  So we introduce a five-argument predicate $sat\_cl5$,
which is going to be delayed.
It defines the set of the lists from $L_{sat\_cl}$ of length greater than 1;
however a list $[Pol1\mydash {\it Var}1, Pol2\mydash {\it Var}2\, |\, {\it Pairs}]$ is
represented as the five arguments
of  $sat\_cl5$.
The intention is to delay selecting $sat\_cl5$ until its first or third
argument is bound (is not a variable).
The specifications for correctness and completeness for $sat\_cl5$ is as
follows.  (As we soon will need another predicate with the same semantics,
the specifications describe both predicates).
{\sloppy
    \[
    \begin{array}{l}
    S_{sat\_cl5}  = \left\{\, \left.
          \begin{array}[c]{@{}l@{\:}}
           sat\_cl5(v_1,p_1,v_2,p_2,s), \\
           sat\_cl5a(v_1,p_1,v_2,p_2,s)
          \end{array}
          \right|
          [p_1\mydash v_1,p_2\mydash v_2|s]\in L_1
    \right\},
    \\[2ex]  %
    S_{sat\_cl5}^0  = \left\{\, \left.
          \begin{array}[c]{@{}l@{\:}}
           sat\_cl5(v_1,p_1,v_2,p_2,s), \\
           sat\_cl5a(v_1,p_1,v_2,p_2,s)
          \end{array}
          \right|
          [p_1\mydash v_1,p_2\mydash v_2|s]\in L_1^0
    \right\}.
    \end{array}
    \]
}%
Now the specifications for the whole program $P_2$ are
\[
\begin{array}{l}
  S_2 = S_1\cup S_=\cup S_{sat\_cl3}\cup S_{sat\_cl5}\ \
      \mbox{(for correctness), and} \\
  S_2^0 = S_1^0\cup S_=\cup S_{sat\_cl3}^0\cup S_{sat\_cl5}^0\ \
      \mbox{(for completeness).}
\end{array}
\]

The following rule completes the definition of $sat\_cl3$.
    \begin{equation}
      \begin{array}l
         sat\_cl3([Pol2\mydash {\it Var}2 | {\it Pairs}], {\it Var}1, Pol1) \gets 
         \\
          \qquad\qquad   sat\_cl5({\it Var}1, Pol1, {\it Var}2, Pol2, {\it Pairs}). 
      \end{array}
     \label{satcl32}
    \end{equation}
To check SC, let $S=S_2$, $L=L_1$  or $S=S_2^0$, $L=L_1^0$.
Then for each ground instance of (\ref{satcl32}), the body is in $S$
iff the head is in $S$
(as
$sat\_cl5(v_1,p_1,v_2,p_2,s)\in S$ iff
$[p_1\mydash v_1,p_2\mydash v_2 | s]\in L$ iff
$sat\_cl3( [p_2\mydash v_2 | s], v_1,p_1)\in S$).
Hence SC1 holds for (\ref{satcl32}), and 
each $sat\_cl3( [p_2\mydash v_2 | s], v_1,p_1)\in S_2^0$ 
is covered by (\ref{satcl32}). 
Thus each 
$sat\_cl3(s',v_1,p_1 )\in S_2^0$ is covered by
(\ref{satcl31}) or (\ref{satcl32}).%
\footnote{%
  Notice that some atoms of the form $sat\_cl(s), sat\_cl3(s,v,p)$ from
  $S_2\setminus S_2^0$  are not covered 
  (e.g. when $s = [a, true\mydash true]$);
  this is a reason why the program is not complete w.r.t.\  $S_2$
  (and  $sat\_cl$ in $P_2$ defines a proper subset of $L_1$).
}

In evaluating  $sat\_cl5$,  we want to treat the bound variable
(the first or the third argument)
in a special way. 
So we make it the first argument of a new predicate
$sat\_cl5a$, with the same declarative semantics as $sat\_cl5$.
\enlargethispage*{1.7ex}
    \begin{eqnarray}
    &&
    sat\_cl5({\it Var}1, Pol1, {\it Var}2, Pol2, Pairs) \gets \nonumber
    \\&&
    \qquad
        sat\_cl5a({\it Var}1, Pol1, {\it Var}2, Pol2, Pairs).
    \label{satcl51}
    \\
    &&
    sat\_cl5({\it Var}1, Pol1, {\it Var}2, Pol2, Pairs) \gets \nonumber
    \\
    &&\qquad
        sat\_cl5a({\it Var}2, Pol2, {\it Var}1, Pol1, Pairs).
    \label{satcl52}
    \end{eqnarray}
Note that 
$[p_1\mydash v_1,p_2\mydash v_2|s]\in L_1$ iff 
$[p_2\mydash v_2,p_1\mydash v_1|s]\in L_1$; the same for $L_1^0$.
It immediately follows that SC are satisfied.
Moreover, SC2 is satisfied by each of the two rules alone.
So each of them is sufficient to define  $sat\_cl5$.
(Formally, the program without (\ref{satcl51}) or without (\ref{satcl52})
remains semi-complete.) 
   The control will choose the one of them which results in invoking $sat\_cl5a$
   with its first argument bound.

To build a procedure $sat\_cl5a$ we have to provide rules
which cover each atom $A= sat\_cl5a(v_1, p_1, v_2, p_2, s)\in S_2^0$.
  Note that $A\in S_2^0$ iff
$[p_1\mydash v_1,p_2\mydash v_2|s]\in L_1^0$ iff
$p_1=v_1$ or $[p_2\mydash v_2|s]\in L_1^0$ iff
$p_1=v_1$ or $sat\_cl3(s,v_2,p_2)\in S_2^0$.
So two rules follow
{\sloppy
  \begin{eqnarray}
  &&\hspace{-2em}
  sat\_cl5a({\it Var}1, Pol1, {\it Var}2, Pol2, {\it Pairs}) \gets {\it Var}1 = Pol1.
  \label{satcl5a1}
  \\
  &&\hspace{-2em}
  sat\_cl5a({\it Var}1, Pol1, {\it Var}2, Pol2, {\it Pairs}) \gets sat\_cl3( {\it Pairs}, {\it Var}2, Pol2).
  \label{satcl5a2}
  \end{eqnarray}
}%
The first one covers $A$ when $p_1=v_1$, the second when
$[p_2\mydash v_2|s]\in L_1^0$.
Thus SC2 holds for each atom $sat\_cl5a(\ldots)\in S_2^0$.
To check SC1,
consider a ground instance of (\ref{satcl5a1}), with the body atom in $S_2$.
So it is of the form
\[
sat\_cl5a(p, p, v_2, p_2, s)
 \gets p= p.
\]
As
$[p\mydash p,p_2\mydash v_2|s]$ is in $L_1$, the head of the rule is in
$S_2$.
Take now a ground instance 
$$
sat\_cl5a(v_1, p_1, v_2, p_2, s) \gets sat\_cl3( s, v_2, p_2).
$$
of (\ref{satcl5a2}), with the body atom in $S_2$.
Then $[p_2\mydash v_2|s]\in L_1$,
hence $[p_1\mydash v_1,p_2\mydash v_2|s]\in L_1$,
and thus $sat\_cl5a(v_1, p_1, v_2, p_2, s)\in S_2$.

From a declarative point of view, our program is ready.
The logic program $P_2$ consists of rules 
(\ref{satcnf1})\,--\,(\ref{satcl5a2}).
As it satisfies SC,
it is correct  w.r.t.\ $S_2$ and semi-complete w.r.t.~$S_2^0$.

\subsection{Avoiding floundering.}
\label{sec:nonfloundering}
As described above, program $P_2$ is intended to be executed with delaying 
certain calls of procedure $sat\_cl5$. 
A $sat\_cl5$-atom will not be selected when the first and the third
argument of $sat\_cl5$ are both variables
(SICStus block declaration {\tt sat\_cl5(-,?,-,?,?)}).
So the program may flounder;
a nonempty query with no selected atom may appear in a computation.%
\footnote{%
  For instance, a query ${\it sat\_c n f}([[{\tt true}\mydash X,{\tt false}\mydash Y]])$
  would lead to a query
   consisting of a single atom
  $sat\_cl5(X,{\tt true},Y,{\tt false},[\,])$,
  which is never selected.
  On the other hand, a query
  ${\it sat\_c n f}([[{\tt true}\mydash X,{\tt false}\mydash Y],[{\tt false}\mydash X]])$
  would lead to selecting 
  $sat\_cl3([\,],X,{\tt false})$, binding $X$ to ${\tt false}$, and then 
  $sat\_cl5({\tt false},{\tt true},Y,{\tt false},[\,])$ is not delayed.
}
Floundering is a kind of pruning of SLD-trees, and may cause incompleteness.
We now augment $P_2$ to avoid floundering.
Then we prove that the obtained program $P_3$ actually does not flounder
(under the intended selection rule, for the intended initial queries). 

 Consider an SLD-derivation starting from a query
 $Q_0 ={\it sat\_c n f}(u)$, where $u$ is a CNF formula.
   Speaking informally, in each query in the derivation each variable is
   bound to some variable of $u$.
 So to avoid floundering we take care that each variable of
 $Q_0$ is eventually bound to a constant.
This may be seen as applying the test-and-generate paradigm.
We add a top level two-argument predicate $sat$. 
It defines the Cartesian product of 
two sets:
(i)~the relation (set) defined by ${\it sat\_c n f}$, and
(ii)~the set of lists of truth values (i.e.\ of {\tt true} or {\tt false}).
The intention is to use initial queries of the form
\begin{equation}
  \begin{array}{@{}c@{}}
  sat(u,l),   \mbox{ where}
  \begin{array}[t]{l@{}}
    u \mbox{ is a (representation of a) propositional formula in CNF}
    \\
    l \mbox{ is the list of the variables in }u.
  \end{array}
  \end{array}
  \label{initial.query}
\end{equation}
For such queries each variable of $u$ will be eventually bound to
{\tt true} or {\tt false}, thus floundering is avoided.

More formally, the specification for the new predicate, respectively for its
correctness and completeness, is
\[
\begin{array}{c}
S_{\it sat} = \{\, sat(u,t) \mid u\in L_2,\ t\in{\it T F L} \,\},
\\
S_{\it sat}^0 = \{\, sat(u,t) \mid u\in L_2^0,\ t\in{\it T F L} \,\},
\end{array}
\vspace{-1ex}
\]
where
\[
 {\it T F L} =
 \{\, [\seq t] \mid n\geq0,\
                     t_i={\tt true} \mbox{ or } t_i={\tt false} 
                     \mbox{ for } i=1,\ldots,n  \,\}.
\]

Now the approximate specification for the extended program consists of
\[
    S_3^0 = S_2^0\cup S_{\it sat}^0 \cup S_{\it t f},
    \qquad
     \qquad\qquad
    S_3 = S_2\cup S_{\it sat} \cup S_{\it t f},
\vspace{-1.5ex plus 1ex}
\]
where
\vspace{-1.5ex plus 1ex}
\[
      S_{\it t f} =
      \{\,    {\it t f list}(t) \mid t\in {\it T F L}  \,\}
      \cup {} 
      \{\, {\it t f}({\tt true}),   {\it t f}({\tt false}) \,\}  
\]
describes two auxiliary predicates.
The three new predicates are defined in a rather obvious way, following \cite{howe.king.tcs-shorter}:
{ \newcommand{\mynegskip}{\hspace*{-3em}}
    \begin{eqnarray}
    &&
    \mynegskip
    sat(Clauses, {\it Vars}) \gets
        {\it sat\_c n f}(Clauses), {\it t f list}({\it {\it Var}s}).   
    \label{sat}
    \\
    &&
   \mynegskip
    {\it t f list}([\,]). 
    \label{tflist1}
    \\&&
   \mynegskip
    {\it t f list}([{\it Var} | {\it Vars}]) \gets
        {\it t f list}({\it Vars}), {\it t f}({\it Var}).
    \\&&
   \mynegskip
    {\it t f}({\tt true}). \\&&     \mynegskip  {\it t f}({\tt false}).
    \label{tf2}
    \end{eqnarray}
}%
We leave for the reader checking of SC (which is trivial for $sat$ and
${\it t f}$, and rather simple for ${\it t f list}$).
\label{final.program}
This completes our construction.
The final logic program $P_3$ consists of rules (\ref{sat})\,--\,(\ref{tf2})
and of the rules of $P_2$, i.e.\
(\ref{satcnf1})\,--\,(\ref{satcl5a2}).%
\footnote{%
\newcommand{\mysymbol}{\raisebox{1pt}{\tiny$\triangleright$}}%
\newcounter{rememberequation}%
\setcounter{rememberequation}{\theequation}%
\setcounter{equation}{6}%
\label{wholeP2}%
    To present the whole program $P_3$ in one place, here is $P_2$.
    The rules of $P_3$,
    except of those marked~\mysymbol\ here,
    will appear unchanged in the final Prolog program.
    \begin{align}
&
      {\it sat\_c n f}([\,]).   \\
&
      {\it sat\_c n f}([Clause | Clauses]) \gets
        sat\_cl(Clause),
        {\it sat\_c n f}(Clauses). 
      \\
&
    sat\_cl([Pol\mydash {\it Var} | {\it Pairs}]) \gets sat\_cl3({\it Pairs}, {\it Var}, Pol). 
      \\
\mysymbol\ \ &
        {=}(X,X).
      \\
&
        sat\_cl3([\,], {\it Var}, Pol) \gets {\it Var} = Pol. 
      \\
&
      sat\_cl3([Pol2\mydash {\it Var}2 | {\it Pairs}], {\it Var}1, Pol1) \gets 
         sat\_cl5({\it Var}1, Pol1, {\it Var}2, Pol2, {\it Pairs}). 
    \\
\mysymbol\ \ &
        sat\_cl5({\it Var}1, Pol1, {\it Var}2, Pol2, Pairs) \gets
            sat\_cl5a({\it Var}1, Pol1, {\it Var}2, Pol2, Pairs).
        \\
\mysymbol\ \ &
         sat\_cl5({\it Var}1, Pol1, {\it Var}2, Pol2, Pairs) \gets 
            sat\_cl5a({\it Var}2, Pol2, {\it Var}1, Pol1, Pairs).
      \\
\mysymbol\ \ &
      sat\_cl5a({\it Var}1, Pol1, {\it Var}2, Pol2, {\it Pairs}) \gets {\it Var}1 = Pol1.
      \\
\mysymbol\ \ &
      sat\_cl5a({\it Var}1, Pol1, {\it Var}2, Pol2, {\it Pairs}) \gets sat\_cl3( {\it Pairs}, {\it Var}2, Pol2).
    \end{align}
\vspace*{-2\belowdisplayskip}
\setcounter{equation}{\therememberequation}%
} %
The program is correct w.r.t.\ $S_3$ and semi-complete w.r.t.\  $S_3^0$.

\paragraph{Non-floundering -- proof.}
Analysis tools, like that of \cite{king.non-suspension2008},
can be used to demonstrate non-floundering of $P_3$, 
for initial queries of the form (\ref{initial.query}),
under a partial selection rule as described above \cite{king.private2012}.
To make the paper self-contained, we present an explicit proof.
We first show that a certain class of programs and queries does not flounder
under a certain class of partial selection rules.

\begin{theorem}[non-floundering]
\label{th:test-generate}
Consider programs $P'_1,P'_2$. 
Let $F_i$ be the set of the predicate symbols occurring in $P'_i$ (for $i=1,2$),
and assume that  $F_1\cap F_2 =\emptyset$.
Let us call an atom $A$ an $F_i$-atom if the predicate symbol of $A$ is in
$F_i$. 

Assume that
each variable occurring in a rule of $P'_1$ occurs in the head of the rule, 
and each variable occurring in a rule of $P'_2$ occurs in the body of the rule.
Consider a partial selection rule $\cal R$ under which each floundered query
consists solely of non-ground $F_1$-atoms.
Let $Q$ be a query such that each variable of $Q$ occurs in
an $F_2$-atom of~$Q$.

Then no derivation of $P'_1\cup P'_2$ for $Q$ via $\cal R$ flounders.
\end{theorem}

We now apply the theorem to our case.
Note that the premises of the theorem are satisfied by the
program $P_4= P_3\setminus \{(\ref{sat})\}$ divided into
$P'_{1}=P_2$ and $P'_{2}=\{(\ref{tflist1}),\ldots,(\ref{tf2})\}$,
and by the selection rule $\cal R$ intended for $P_3$, and described previously.
(Any floundered query consists of non-ground $sat\_cl5$-atoms,
  $F_1$ contains the predicate symbols of $P_2$, and
 $F_2 = \{ {\it t f list}, {\it t f} \}$.)
Consider a (finite or infinite)
 SLD-derivation $Q_0, Q_1,\ldots$ of $P_3$, where $Q_0$ is of the
form (\ref{initial.query}).  Then $Q_1,\ldots$ is a derivation of $P_4$, and
$Q_1={\it sat\_c n f}({u}),{\it t f list}(l)$ satisfies the conditions of the
theorem.
Thus $P_3$ does not flounder under $\cal R$ for the intended initial queries.

\begin{proof}[Proof \rm (of Theorem \ref{th:test-generate})]
Consider a (finite or infinite)
derivation $Q\mathop= Q_0, Q_1,\ldots$ of $P'_1\cup P'_2$ via $\cal R$.
We show that (independently from the selection rule)
\begin{equation}
\label{property-queryvars}
  \parbox{.66\textwidth}
  {
  if a variable $X$ occurs in an $F_1$-atom $A$ in $Q_i$  ($i>0$)
  then $X$ occurs in an $F_2$-atom $B$ in $Q_i$.
  }
\end{equation}
Hence no $Q_i$ consists of non-ground $F_1$-atoms, and thus floundering does
not occur. 

As each variable of a clause of $P'_1$ occurs in the head, we have:
\begin{equation}
\label{property-stepvars}
  \parbox{.86\textwidth}
         {
           Assume that in a resolution step an $F_1$-atom $A$ in a query 
           is replaced by $(\seq[k]A)\theta$ (where $\theta$ is an mgu of $A$
           and the head $H$ of a rule with body $\seq[k]A$).

  If
           a variable $X$ occurs in $(\seq[k]A)\theta$ then $X$ occurs in
           $A\theta$. 
         }
\end{equation}
(As $X$ occurs in some $V\theta$, where $V$ occurs both 
in the body  $\seq[k]A$ and in the head $H$;
thus $V\theta$ occurs both in $H\theta$ and $A\theta$.)

Similarly, due to each variable of a clause of $P'_2$ occurring in the body,
we have:
\begin{equation}
\label{property-stepvars2}
  \parbox{.86\textwidth}
         {
           Assume that $A$ is an $F_2$-atom and, apart from this,
           $A,\seq[k]A,\theta$ are as in 
           (\ref{property-stepvars}).
           If a variable $X$ occurs in $A\theta$ then  $X$ occurs in
           $(\seq[k]A)\theta$. 
         }
\end{equation}
(The explanation is similar.)
The proof of  (\ref{property-queryvars}) is  by induction.
  (\ref{property-queryvars}) holds for $Q_0$.
Assume it holds for a $Q_i$, $i>0$.  The next query
$Q_{i+1}$ is obtained from $Q_i$ by applying an mgu $\theta$
after a replacement of an $F_1$-atom by some $F_1$-atoms $\seq[k]A$, 
or a replacement of an $F_2$-atom by some $F_2$-atoms.
Assume that a variable $Y$ occurs in an $F_1$-atom of $Q_{i+1}$.
Then either by (\ref{property-stepvars}) $Y$ occurs in $A\theta$ 
for the selected $F_1$-atom $A$ of $Q_i$,
or $Y$ occurs in an atom $A\theta$ of $Q_{i+1}$, where $A$ is
a (not selected) $F_1$-atom of $Q_i$.   
In both cases, $Y$ occurs in $X\theta$, for some $X$ occurring in $A$;
hence, by the inductive assumption, $X$ occurs in some $F_2$-atom $B$ of
$Q_i$, thus $Y$ occurs in $B\theta$.  
Either $B\theta$ is an $F_2$-atom of $Q_{i+1}$, 
or by (\ref{property-stepvars2}) $Y$ occurs in an $F_2$-atom of $Q_{i+1}$.
\end{proof}

The author is not aware of any existing method applicable in our case.
The sufficient condition for non-floundering of
\cite[Theorem 3.5]{SmausHK98.PLILP98}, a generalization of that of 
\cite{AptL95.delays}, is based on 
assigning types and directions (input or output) to argument positions.
For the condition to hold for program $P_3$,
the first and the third argument of $sat\_cl5$ have to be input.
Consequently, some other arguments have to be input, including
the first argument of ${\it sat}$.
But then the condition for initial queries of the form (\ref{initial.query})
is violated
(for any type assignment under which the condition holds for $P_3$).
It seems however that
the condition may be used to construct a proof similar to ours (p.\,18), by 
applying it to query $Q_1$ and program  $P_1'\cup P_2'$.

\subsection{Completeness and termination of $P_3$.}
We showed that  $P_3$ is correct and semi-complete.
To establish its completeness we show that it is recurrent.
Consider a level mapping
\[
\begin{array}{l}
  |sat(t,u) | = \max\left(\, 3|t|,\,list size(u)\, \right) + 2,    \\
  | {\it sat\_c n f}(t) | = 3|t|+1, \\
    | sat\_cl(t)| = 3|t|+1,  \\
| sat\_cl3(t,u_1,u_2)| = 3|t|+1,  \\
\end{array}
\quad
\begin{array}{l}
| sat\_cl5(u_1,u_2,u_3,u_4,t)| = 3|t|+3,  \\
| sat\_cl5a(u_1,u_2,u_3,u_4,t)| = 3|t|+2,  \\
  | {\it t f list}(u) |      = list size(u), \\
 |t=u| = | {\it t f}(t) | = 0,
\end{array}
\]
where $t,u,u_1,u_2,u_3,u_4$ are arbitrary ground terms,
$|t|$ is as in Section \ref{sec:logicProgram1}
and $list size$ is defined by
$list size([h|t])=list size(t)+1$ and 
$list size(f(\seq t))=0$ for any $f$ which is not $[\ | \ ]$.
For example consider (\ref{satcl}).  For any its ground instance
$sat\_cl([p\mydash v|t])\gets sat\_cl3(t,v,p)$,
the level mapping of the head is $3|t|+4$, while that of the body atom is
$3|t|+1$.  We leave to the reader further details of the proof that $P_3$ is
recurrent. 
As it is semi-complete, it is complete by  Corollary \ref{cor:completeness:recurrent}.

As an additional corollary we obtain a non-declarative property --
termination of $P_3$ under any selection
rule for the intended initial queries.
 Consider a query
$Q=sat(t,t')$, where 
$t$ is a list of lists of elements of the form  $s\mydash s'$,
and $t'$ is a list.  Each intended query to the program is of this
form.  $Q$ is bounded (for each its ground instance $Q\theta$, $|Q\theta|=|Q|$).
As $P_3$ is recurrent, each SLD-tree for $P_3$ and $Q$ is finite.
\subsection{Occur-check}
\label{sec:occur-check}
The implementation of unification in Prolog is unsound, due to lack of the
occur-check.  As a result, Prolog may produce answers which, according to the
theory, are not answers of the program.
Here we show that 
the occur-check in the unification algorithm is unnecessary
for program $P_3$, initial queries of the form (\ref{initial.query}),
and any selection rule.
So the correctness of $P_3$ w.r.t.\ $S_3$ is preserved when the program (with
such queries) is executed by Prolog.

First, each rule head of $P_3$, except for that of (\ref{ruleeq}), is
linear.
So the occur-check is unnecessary to correctly unify such a head with any atom.
More formally, the corresponding equation is NSTO (not subject to occur-check)
\cite[Lemma 7.5]{Apt-Prolog}.

It remains to consider rule  (\ref{ruleeq}), which is 
${=}(X,X)$.
Let $Q_0,Q_1,\ldots$ be an SLD-derivation for $P_3$
starting from a query $Q_0$ of the form  (\ref{initial.query}).
By simple induction on $i$ we obtain the following facts.  
In each subterm $t\mydash v$ of any query $Q_i$,
    $t$ is a constant (it is {\tt true} or {\tt false}).  
Also,
in each $sat\_cl3$-atom in $Q_i$ the third argument is a constant
and the second and the fourth arguments of any $sat\_cl5$- or $sat\_cl5a$-atom
are constants.

Note now that each  $=$-atom $v{=}p$ appearing in a query 
has been introduced as the result of a resolution step employing 
rule (\ref{satcl31}) or (\ref{satcl5a1}).
Hence $p$ is a constant, and the occur-check in resolving $v{=}p$  with rule
(\ref{ruleeq}) is unnecessary
-- the Martelli-Montanari unification algorithm eventually unifies $v$ with $p$.%
\footnote{%
    Formally, this follows from analysis of the behaviour of the algorithm.
    Another proof can be obtained
    by Lemma 7.5 of \cite{Apt-Prolog}).

    Also, having proved that $p$ is ground in any $v{=}p$ in any query,
    we obtain the occur-check freedom by
    Occur-check Freedom 1 theorem of
    \cite{AptL95.delays} (with the second position of $=$ treated as input,
    and all the other argument positions as output).
} %
To summarize, $P_3$ can be safely executed under the unsound unification of
Prolog, lacking the occur-check.

A comment is due that the notion of a NSTO program (with a query) of 
\cite{Apt-Prolog}
is irrelevant here, as it guarantees occur-check 
freedom only for LD-derivations.
The corresponding notion of \cite{Deransart.Maluszynski93}
deals with arbitrary selection rules, but $P_3$ with $Q_0$ does
not satisfy the first two sufficient conditions for NSTO given in that work
(Proposition 8.3, Theorem 8.5).
The third sufficient condition involves a substantial transformation of the
program, and seems substantially more 
complicated than the proof presented here.
Also the method of \cite{AptL95.delays} and its generalizations of
\cite{SmausHK98.PLILP98,SmausHK98.LOPSTR98} are inapplicable in our case.
The methods are based on modes -- classifying argument positions as input or
output.  Occur-check freedom is guaranteed for so called nicely moded
programs and queries.  In our case, $P_3$ with $Q_0$ is not nicely moded
under any moding.%
\footnote{%
   At least one position of $=$ must be output, hence some positions of 
   $sat\_cl3$, $sat\_cl5a$,  $sat\_cl5$, and $sat\_cl$ must be output.
   Thus the first position of ${\it sat\_c n f}$  is output; this requires
   the (term representing a) clause in the initial query to be linear.
}

\subsection{On correctness-preserving transformations}
\label{sec:transformations}
Here we discuss inapplicability of the paradigm of semantics preserving
program transformations to our example.

Let us first point out that our example shows usefulness of approximate
specifications.  In Section \ref{sec:logicProgram1}
we showed that the problem statement naturally leads to an approximate
specification.  It is unnecessary to know the exact relations defined by the
program; what matters is its full correctness w.r.t.\ the approximate
specification   (i.e.\ correctness w.r.t.\ the specification for correctness
and completeness w.r.t.\ that for completeness).
Note that the exact semantics of programs $P_2$ and $P_3$ is not obvious, and
describing it may be rather cumbersome.  
For instance, $sat\_cl$ defines (both in $P_2$ and $P_3$) the set of terms of
the form $[t\mydash t]$, or
$[t_1\mydash u_1,\ldots,t_n\mydash u_n|s]$  ($n>1$), where  $t_i=u_i$
for some $i$. 
Moreover, the semantics of (the common predicates in) the consecutive versions 
$P_1, P_2$ of the program differ.  
For instance, $A = sat\_cl([{\tt true}\mydash{\tt true}|1])\in \M_{P_1}$, 
but $A\not\in \M_{P_2}$.
What does not change, is the correctness
and completeness w.r.t.\ a fixed approximate specification (of the common
predicates of $P_1, P_2, P_3$).

This shows a weakness of the paradigm of program
development by seman\-tics-preserving transformations
\cite[and the references therein]{PettorossiPS10shorter}.
Constructing $P_3$ by semantics-preserving transformations seems completely
unfeasible. 
The exact semantics of (the common predicates in) the final
program depends on particular design decisions made 
at the development steps and cannot be known in advance.
Thus our example suggests that a more general paradigm of program
transformations is worth introducing -- transformations which
preserve correctness and completeness w.r.t.\ an approximate specification.

\section{Pruning and completeness}
\label{sec:compl-pruning}
In the previous section we constructed a logic program $P_3$,
which is correct and complete w.r.t.\ the respective specifications
and, for the intended queries, terminates, is occur-check free, and does not
flounder under the intended selection rule.
The final Prolog program will be obtained from $P_3$ by adding control;
this includes pruning.
This section presents a way of proving that pruning does not violate
program completeness.

Pruning means removing some parts of SLD-trees.
In Prolog, pruning can be performed by means of the cut, or devices like 
{\tt once}/1 or the if-then-else construct.
Pruning preserves correctness of a program;
it also preserves termination,
occur-check freedom, and non-floundering.
However it may violate program completeness.
We present a method \cite{drabent.tocl16}
of showing that completeness is actually preserved. 
The presentation follows  \cite{drabent.tocl16}.
This method is based on a rather abstract view of pruning; 
a (more complicated) approach
directly referring to the cut is presented in \cite{drabent.faoc16}.

\subsection{Pruned SLD-trees}

\begin{minipage}{.55\textwidth}
\vspace{1.3ex}
    We will view pruning of an SLD-tree as applying only certain rules
    while constructing the children of a node.
    To formalize this, we introduce subsets $\seq\Pi$ of a given program $P$.  
    The intention is that for each node the rules of one $\Pi_i$ are used.
  The diagram \cite{drabent.lopstr14}
  compares selection of atom $A$ in an SLD-tree with selection of
  $A$ and $\Pi_i$ in a pruned tree.
\end{minipage}
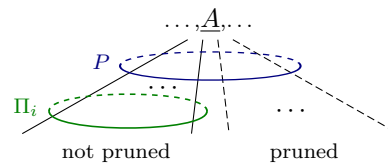
\begin{minipage}{.45\textwidth}

\newcommand{\myellipse}{%
    \begin{pgfpicture}{-1.5cm}{-.5cm}{1.5cm}{.5cm}
      \pgfsetxvec{\pgfpoint{.5cm}{0cm}}
      \pgfsetyvec{\pgfpoint{0cm}{.5cm}}
      \pgfsetlinewidth{1pt}
      \pgfmoveto{\pgfxy(-3,0)}
      \pgfcurveto{\pgfxy(-3,-1)}{\pgfxy(3,-1)}{\pgfxy(3,0)}
      \pgfstroke
      \pgfmoveto{\pgfxy(3,0)}
      \pgfsetdash{{3pt}{3pt}}{0pt}
      \pgfcurveto{\pgfxy(3,1)}{\pgfxy(-3,1)}{\pgfxy(-3,0)}
      \pgfstroke
    \end{pgfpicture}
}
  \newcommand{\dblue}{\color[rgb]{0,0,.5}}
  \newcommand{\dgreen}{\color[rgb]{0,0.5,0}}



\newcommand{\mytreepicture}[1][{}]
{%
\begin{pgfpicture}{-2.5cm}{-.5cm}{2.5cm}{1.7cm}
      \pgfsetxvec{\pgfpoint{.5cm}{0cm}}
      \pgfsetyvec{\pgfpoint{0cm}{.5cm}}
  \pgfputat{\pgfxy(0,3)}{\pgfbox[center,center]
    {{\footnotesize\ldots,}$\underline A${\footnotesize,\ldots}}
  }
  \pgfline{\pgfxy(-0.6,2.5)}{\pgfxy(-5,0)}
  \pgfline{\pgfxy(-0.2,2.5)}{\pgfxy(-.5,0)}
  \begin{pgfscope}
    \pgfsetdash{{3pt}{1.5pt}}{0pt}
    \pgfline{\pgfxy(0.2,2.5)}{\pgfxy(0.5,0)}
    \pgfline{\pgfxy(0.6,2.5)}{\pgfxy(5,0)}
  \end{pgfscope}


  \pgfputat{\pgfxy(-4.9,.7)}
           {\pgfbox[center,center]{\mbox{\small\footnotesize$\dgreen\Pi_{#1}$}}}
  \pgfputat{\pgfxy(-1.1,1)}{
    \pgfbox[center,center]{\dgreen
      \begin{pgfmagnify}{.7}{.6}
          \myellipse
      \end{pgfmagnify}
      }
  }
  \pgfputat{\pgfxy(-1.2,1.2)}{{\pgfbox[center,center]{$\cdots$}}}
  \pgfputat{\pgfxy(+2.2,0.6)}{{\pgfbox[center,center]{$\cdots$}}}

  \pgfputat{\pgfxy(-2.9,1.9)}
           {\pgfbox[center,center]{\mbox{\small\footnotesize$\dblue P$}}}
  \pgfputat{\pgfxy(-2.4,1.3)}{\dblue
      \begin{pgfmagnify}{.8}{.5}
          \myellipse
      \end{pgfmagnify}
  }      
  \pgfputat{\pgfxy(2.5,-.5)}
           {\pgfbox[center,center]{\mbox{\small\footnotesize pruned}}}  
  \pgfputat{\pgfxy(-2.5,-.5)}
           {\pgfbox[center,center]{\mbox{\small\footnotesize not pruned}}}  
\end{pgfpicture}
} 

  %
\mbox{}\hfill
\mbox{\mytreepicture[i]}
\hspace*{-.5em}
\end{minipage}

\begin{definition}
\nopagebreak
Given programs $\seq{\Pi}$ ($n>0$),
a {\bf c-selection rule} is a function%
\
assigning to a query $Q'$ an atom $A$
in $Q'$ and one of the programs $\emptyset,\seq{\Pi}$.

A {\em pruned SLD-tree}, or {\bf csSLD-tree} (cs for clause selection), for a
query 
$Q$ and programs $\seq\Pi$, via a c-selection rule $R$,
is constructed as an SLD-tree, but
for each node its children are constructed using the program selected by the
c-selection rule.
An {\em answer} of the csSLD-tree is the answer of a successful derivation
which is a branch of the tree.

A csSLD-tree $T$ with root $Q$ is {\bf complete} w.r.t.\ a specification $S$
if, for any ground  $Q\theta$,
$S\models Q\theta$ implies that $Q\theta$ is an instance of an answer of $T$.

\end{definition}
Informally, such a complete tree produces all the answers for $Q$ required by $S$.
A notion similar to c-selection rule was used by 
\citeN{PedreschiRS02.TPLP.terminating}.

In Prolog with delays it is possible that no atom is selected in a non-empty
query.  This can be modelled as a c-selection rule which for such query
selects program $\emptyset$.

\subsection{Reasoning about completeness of csSLD-trees}

Consider programs $P,\seq\Pi$ and specifications $S,\seq S$, such that 
$P\supseteq\bigcup_{i=1}^n\Pi_i$ and $S=\bigcup_{i=1}^n S_i$.
The intention is that each $S_i$ describes which answers are to be produced by
using $\Pi_i$ in the first resolution step.
We will call $\seq {\Pi}$, $\seq S$ a {\bf split} (of $P$ and $S$).

\begin{definition}
\label{def:suitable}
Let  $\S =\seq {\Pi}$, $\seq S$ be a split, and $S=\bigcup S_i$.
Specification $S_i$ is {\bf suitable} for an atom $A$
w.r.t.\ \S
when $ground(A)\cap S \subseteq S_i$.
(In other words, when no instance of $A$ is in $S\setminus S_i$.)
We also say that a program $\Pi_i$ is {\bf suitable} for $A$ w.r.t.\ \S
when $S_i$ is.

A c-selection rule is {\bf compatible} with \S if for each non-empty query
$Q'$ it selects an atom $A$
and a program $\Pi$, such that 

\quad
-- $\Pi\in\{\seq\Pi\}$ is suitable for $A$ w.r.t.\ \S, or

\quad
-- none of $\seq\Pi$ is suitable for $A$ w.r.t.\ \S and  $\Pi=\emptyset$
(so $Q'$ is a leaf).

A csSLD-tree for $\seq {\Pi}$ via a c-selection rule compatible with \S
is said to be {\bf weakly compatible} with \S.
The tree is {\bf compatible} with \S when 
for each its nonempty node some $\Pi_i$ is selected.
\end{definition}

  The intuition is that 
  when $\Pi_i$ is suitable for $A$
  then $S_i$ is a fragment of $S$ sufficient to deal with $A$.
  It describes all the answers for query $A$ required by~$S$.
For instance, consider a specification  
$S = \{\, p(t)\mid t\in L_1\,\}$ (cf.\,(\ref{L1L2}))
and a split $\S = \Pi_1,\Pi_2,S'_1,S'_2$,
where $S'_1 = \{\,p([t\mydash t|s]) \mid t,s\in\HU \,\}$ and $S'_2=S$
(we skip the details of $\Pi_1,\Pi_2$, note that $S'_1\subseteq S$).
Then $S'_1$ (and $\Pi_1$) is suitable for an atom $A = p( [X\mydash X|T])$,
and is not suitable for $p( [X\mydash Y|T])$.

In a compatible tree, $\Pi_i$ is selected together with $A$.
If now each atom of $S_i$ is covered by  $\Pi_i$ w.r.t.\ $S$ then using for
$A$ only the rules of $\Pi_i$ does not destroy completeness.
Formally \cite{drabent.tocl16}:

\vspace{0pt plus 5pt}
\pagebreak[3]
\begin{theorem}
\nopagebreak
\label{th:completeness:pruned}%
    Let $P\supseteq\bigcup_{i=1}^n\Pi_i$ (where $n>0$) be a program, 
    $S=\bigcup_{i=1}^n S_i$ a specification, and 
    $T\!$ a csSLD-tree for $\seq\Pi$.
    If
    \begin{enumerate}[\qquad]
    \item 
    \label{prop:cssld.complete.cond0}
        for each $i=1,\ldots,n$, 
        all the atoms from $S_i$ are covered by $\Pi_i$ w.r.t.\ $S$, and 
    \item
    \label{prop:cssld.complete.cond1}
        $T$ is compatible with $\seq\Pi,\seq S$,
    \item 
    \label{prop:cssld.complete.cond2}
    $T$ is finite or
         $P$ is recurrent  %
    \end{enumerate}
    then $T$ is complete w.r.t.\ $S$.
\end{theorem}

\section{The program with control}
\label{sec:prologprogram}
 In this section we add control to program $P_3$.
As the result we obtain the Prolog program of  \citeN{howe.king.tcs-shorter}.  
(The
 predicate names differ, those in the original program are related
 to its operational semantics.)
 The idea is that $P_3$ with this control implements
 the DPLL algorithm with watched literals and unit propagation.%
\footnote{%
However, removing a clause when a literal in it becomes true is implemented
only when the literal is watched in the clause.
}
The control added to $P_3$ modifies
 the default Prolog selection rule (by means of delays),
and prunes some redundant parts of the search space (by the if-then-else
construct).  So correctness, termination, and occur-check freedom of $P_3$
are preserved 
(as the latter two properties have been proved for any selection rule).  

The first control feature to impose is delaying $sat\_cl5$ until 
its first or third argument is not a variable.  This can be done by a
SICStus block declaration
    \begin{equation}
    \label{block}
      \mbox{\tt:- block sat\_cl5(-, ?, -, ?, ?)}.
    \end{equation}
For the intended initial queries, such delaying does not lead to floundering
(as shown in Section \ref{sec:nonfloundering}). So the completeness of the logic program
is preserved.

The first case of pruning is to use
only one of the two rules (\ref{satcl51}), (\ref{satcl52}),
\par
{\footnotesize
 \[
 \begin{array}{l}
 sat\_cl5({\it Var}1, Pol1, {\it Var}2, Pol2, {\it Pairs}) \gets
 sat\_cl5a({\it Var}1, Pol1, {\it Var}2, Pol2, {\it Pairs}).
 \\
 sat\_cl5({\it Var}1, Pol1, {\it Var}2, Pol2, {\it Pairs}) \gets
 sat\_cl5a({\it Var}2, Pol2, {\it Var}1, Pol1, {\it Pairs}).
 \end{array}
 \]
}%
the one which invokes $sat\_cl5a$ with the first argument bound.
We achieve this by employing the $non var$ built-in and the if-then-else
construct of Prolog:
\begin{equation}
\label{satcl5P}
 \begin{array}{l}
   sat\_cl5({\it Var}1, Pol1, {\it Var}2, Pol2, {\it Pairs}) \gets  
   \\\qquad
   non var({\it Var}1) 
   \begin{array}[t]{cl}
       \to & sat\_cl5a({\it Var}1, Pol1, {\it Var}2, Pol2, {\it Pairs})
       \\
       ; &    sat\_cl5a({\it Var}2, Pol2, {\it Var}1, Pol1,  {\it Pairs}).
   \end{array}
 \end{array}
\end{equation}

The other case of pruning concerns rules (\ref{satcl5a1}), (\ref{satcl5a2}):
\par
{\footnotesize
 \[
 \begin{array}{l}
  sat\_cl5a({\it Var}1, Pol1, {\it Var}2, Pol2, {\it Pairs}) \gets {\it Var}1 = Pol1.
\\
  sat\_cl5a({\it Var}1, Pol1, {\it Var}2, Pol2, {\it Pairs}) \gets sat\_cl3( {\it Pairs}, {\it Var}2, Pol2).
 \end{array}
 \]
}%
Rule (\ref{satcl5a2}) is to be skipped when (\ref{satcl5a1}) succeeds.%
\
(Here we prune further search when clause 
$[{Pol1,}{\it Var}1, Pol2, {\it Var}2\, | {\it Pairs}]$ 
is found true.)
This is done by converting the two rules into
\begin{equation}
\label{satcl5aP}
  \begin{array}{l}
  sat\_cl5a({\it Var}1, Pol1, {\it Var}2, Pol2, {\it Pairs}) \gets
  \\\qquad
   {\it Var}1 \mathop= Pol1 \, \to \, true \ ;  \
   sat\_cl3( {\it Pairs}, {\it Var}2, Pol2).
  \end{array}
\end{equation}

This completes our construction.
The obtained Prolog program consists of declaration (\ref{block}),
the rules of $P_3$ except for those for $sat\_cl5$, $sat\_cl5a$,  and
$=$, i.e.\ 
(\ref{satcnf1})\,--\,(\ref{satcl}), (\ref{satcl31}), (\ref{satcl32}), 
(\ref{sat})\,--\,(\ref{tf2}),
and Prolog rules  (\ref{satcl5P}), (\ref{satcl5aP}).
Rule (\ref{ruleeq}) for $=$ has to be skipped, as it is built in Prolog.%
\footnote{%
      Note that whenever $v=p$ originating from (\ref{satcl5aP}) is selected
      (in a computation starting from an initial query of the form
      (\ref{initial.query})) then $v$ is not a variable 
      (by (\ref{satcl5P}) and (\ref{block}))
      and $p$ is a constant (as shown in Section \ref{sec:occur-check}).
      Thus $=$ in (\ref{satcl5aP}) can be replaced by
      a non-logical built-in {==} of Prolog, 
      as it is done in \cite{howe.king.tcs-shorter}.
}

\paragraph{Completeness of the final program.}
We now prove that completeness is preserved under pruning imposed on $P_3$
by the control described above.  Let us construct a split 
$\S = \Pi_0,\ldots,\Pi_5,S'_0,\ldots,S'_5$
of program $P_3$ and its specification for completeness
$S_3^0$:
\[
\begin{array}{l@{}l}
\Pi_1 = \{\, (\ref{satcl51}) \,\},
& 
S'_1 = S'_2 =  S_3^0\cap\{\, sat\_cl5(t_1,\ldots,t_5)  \mid 
                   t_1,\ldots,t_5\in \HU   \,\}
\\
\Pi_2 = \{\, (\ref{satcl52}) \,\}, 
&
\quad\ 
 = \{\, sat\_cl5(v_1,p_1,v_2,p_2,s)  \mid 
                    [p_1\mydash v_1,p_2\mydash v_2|s]\in L_1^0 \,\},
\\
\Pi_3 = \{\, (\ref{satcl5a1}) \,\},
& 
    S'_3 = \{\, sat\_cl5a(p,p,v_2,p_2,s) \mid 
                     [p\mydash p,p_2\mydash v_2|s]\in L_1^0 \,\},
\\
\Pi_4 = \{\, (\ref{satcl5a2}) \,\},
& 
    S'_4 = \{\, sat\_cl5a(v_1,p_1,v_2,p_2,s) \mid v_1\neq p_1,\
                     [p_2\mydash v_2|s]\in L_1^0\,\},
\\
\Pi_5 = P_3\setminus\bigcup_{i=1}^4\Pi_i,
&
\qquad\
    S'_5 = S_3^0\setminus (S'_1\cup S'_3\cup S'_4).
\end{array}
\]
Note that $S'_5$ does not contain any $sat\_cl5$-atoms or $sat\_cl5a$-atoms,
and that $\bigcup_{i=1}^5 S'_i = S_3^0$.
In particular, $S'_3\cup S'_4$ is the set of all $sat\_cl5a$-atoms from $S_3^0$.
To apply Theorem \ref{th:completeness:pruned}
we need to show that
for $i=1,\ldots,5$ each atom of $S'_i$ is covered by $\Pi_i$ w.r.t.\ $S_3^0$.
A proof of this fact has already been
done; 
it consists of the relevant fragments of the proof that 
each atom $S_3^0$ is covered by $P_3$, spread over Sections
\ref{sec:logicProgram1},~\ref{sec:program}.
In what follows let us consider
a pruned tree $T$ with the root of the form (\ref{initial.query}).  It
can be seen as a csSLD-tree for $\Pi_1,\ldots,\Pi_5$.  The c-selection rule
selects $\Pi_1$ or $\Pi_2$ for  $sat\_cl5$-atoms, 
$\Pi_3$ for $sat\_cl5a$-atoms with the first two arguments unifiable,
$\Pi_4$ for the remaining $sat\_cl5a$-atoms,
and $\Pi_5$ for the atoms with the other predicate symbols.

We now show that the tree $T$ is compatible with \S.
When  $\Pi_i$, where $i\in\{1,2,5\}$,  is selected for a $q$-atom $A$ then
the corresponding specification $S'_i$ contains all the $q$-atoms from $S_3^0$.
So $ground(A)\cap S_3^0\subseteq S'_i$.
It remains to consider $A$ being a $sat\_cl5a$-atom.
Due to the delays implemented by (\ref{block}) and 
pruning implemented in (\ref{satcl5P}), when a $sat\_cl5a$-atom $A$ is
selected in a node of the tree then its first argument is not a variable,
say it is $f(\seq t)$  ($n\geq0$).
In Section \ref{sec:occur-check} we showed that the second argument is a
constant $c$ (which is {\tt true} or {\tt false}).  
So $A = sat\_cl5a(f(\seq t),c,\ldots)$.
Thus either $ground(A) \cap S_3^0\subseteq S'_3$
(when the first two arguments in $A$ are unifiable, hence are equal)
and $\Pi_3$ is selected for $A$, or
$ground(A) \cap S_3^0\subseteq S'_4$ (otherwise)
and $\Pi_4$ is selected for $A$.
Remember that in Section \ref{sec:nonfloundering} we showed 
that in each non-empty query of the tree some atom is selected (there is no
floundering). 
Thus $T$ is compatible with  \S.

Remember also that $P_3$ is recurrent. 
So the premises of Theorem \ref{th:completeness:pruned}
are satisfied, and each pruned tree with the root of the form
(\ref{initial.query}) is complete w.r.t.\ $S_3^0$.
\paragraph{Remarks.}
The Prolog program constructed in this section
is correct w.r.t.\ specification $S_3$ and, for initial queries of the form 
(\ref{initial.query}), it terminates under any selection rule,
does not flounder, and is occur-check free.  These properties follow from
those of logic program $P_3$, proved in Section \ref{sec:program}.
Here we proved that, for queries of this form,
the Prolog program preserves the completeness 
of $P_3$ w.r.t.~$S_3^0$.
In practice, this means that the program will produce
all the answers required by the specification.
Thus, by (\ref{property.program}), the program will find all the solutions to
the SAT problem represented by the query.

\section{Conclusions}

\paragraph{The example.}
This paper presents a construction of a non-toy Prolog program,
together with proofs of its correctness, completeness,
 termination, occur-check freedom, and non-floundering.
The program is the SAT solver of \citeN{howe.king.tcs-shorter}.  
It is seen as a definite clause logic program with added control.
Starting from a formal specification, a 
sequence of definite clause programs, called $P_1,P_2,P_3$,
is constructed hand in hand with proofs of their correctness and completeness 
(Section \ref{sec:logicProgram1},\,\ref{sec:program}).
The construction is guided by the proofs, mainly those of semi-completeness.
This suggests a general approach (Section \ref{sec:construction})
to constructing programs which are provably correct and semi-complete.
The author believes that the approach, possibly treated informally,
should be useful in practice.

Additionally, for the assumed class of initial queries
program $P_3$ is proved to terminate, not flounder, and be occur-check free;
  termination and occur-check freedom hold for arbitrary selection rules,
  and non-floundering -- for a class of selection rules including that to be
  used in the final Prolog program.

The Prolog program is obtained from $P_3$ by adding control --
imposing a specific selection rule and pruning the SLD-trees.
The pruning is done by means of the if-then-else
construct (and can equivalently be done by means of the cut).
Such control preserves correctness; it also preserves 
occur-check freedom, and termination 
(as in this particular case they are independent from the selection rule).
However it may violate completeness due to pruning 
and floundering.  Non-floundering has already been proven for $P_3$.
In Section \ref{sec:prologprogram} 
we prove that the completeness is preserved under pruning.

\paragraph{Proof methods.}
We applied general methods to prove correctness  \cite{Clark79},
completeness \cite{drabent.tocl16}, termination \cite{DBLP:journals/jlp/Bezem93},
and completeness under pruning  \cite{drabent.tocl16}.
The known methods for proving non-floundering and occur-check freedom seem
inapplicable in our case
(cf.\ Sections \ref{sec:occur-check}, \ref{sec:nonfloundering}).
To deal with non-floundering,
we introduced a sufficient condition, suitable for a certain class of programs.
The proof of occur-check freedom is rather {\em ad hoc}.
It should be added that non-floundering and termination can be
determined by means of automatic tools, see
\cite{king.non-suspension2008} for the former,
and e.g.\ \cite[and the references therein]{NguyenSGS11-termination-shorter} 
for the latter.
{\sloppy\par}

The employed proof methods for correctness and completeness were described
and discussed in \cite{drabent.tocl16}.
  The method for program correctness  \cite{Clark79} is simple, natural,
  should be well-known, but is often neglected.
  Instead, more complicated methods are proposed, like that of
  \cite{Apt-Prolog}. 
  Moreover, these methods are not declarative
  (cf.\ footnote \ref{footnote:nondeclarative}, see
  \cite{DBLP:journals/tplp/DrabentM05shorter,drabent.tocl16}
  for further discussion).
  The methods for program completeness, and 
  for completeness of pruned SLD-trees are due to \cite{drabent.tocl16}.
  The approach for completeness introduces a notion of semi-completeness.
  Roughly speaking, semi-completeness and termination imply completeness.
  Semi-completeness alone may be useful -- if a semi-complete program
  terminates then all the answers required by the
  specification have been obtained.
  The presented sufficient condition for semi-completeness
  is a necessary condition for program completeness
  (in a sense made precise in  \cite{drabent.tocl16});
  so semi-completeness may be understood as a necessary step when dealing
  with completeness.

A subject for future work is formalization and automation of the presented
proof methods.  This requires introducing a particular language to
express (a chosen class of) specifications.  The next issue is devising
formal methods of checking the sufficient conditions used here.

In the author's opinion,
the sufficient conditions for correctness and semi{-}com\-pleteness
are simple and natural.  
For correctness one has to show that the rules of
the program produce specified atoms (cf.\ Definition \ref{def:specification})
 out of specified ones.
For semi-completeness -- that each specified atom can be produced by some rule of
the program out of some specified atoms.
The author believes
  that the two proof methods
  and the program construction approach of Section \ref{sec:construction}
  are a practical and useful tool.  They should be applicable,
  possibly at a less formal level, in actual every-day programming, 
   and in teaching logic programming.
The program development case of Sections
\ref{sec:logicProgram1},~\ref{sec:program},
maybe dealt with less formally, can be considered a practical example.
Conversely,
the two proof methods can be understood as a formalization of a usual way in
which competent programmers reason declaratively about their logic programs.

\paragraph{Declarativeness.}
Logic programming could not be considered a declarative programming paradigm
unless there exist
declarative ways of reasoning about program correctness
and completeness.
(We remind that by declarative we mean referring only to logical reading of 
programs, thus abstracting from any operational semantics.)
The methods for proving correctness and semi-completeness
are purely declarative.  
So is the completeness proving method used in this paper; however it is 
applicable to a restricted class of programs.

In general, completeness proofs based on semi-completeness
 may be not declarative, as 
they refer, maybe indirectly, to program termination.
So we have a compromise between declarative and non-declarative reasoning:
declarative proofs of correctness and semi-completeness, and a
non-declarative step from semi-completeness to completeness.
Nevertheless, such compromise 
should be useful in the practice of declarative programming,
as usually termination has to be established anyway.
This is also supported by the fact that
a declarative method for proving completeness \cite{Deransart.Maluszynski93}
results in a proof similar to a proof of semi-completeness together with a
proof of termination (Section \ref{subsec:completeness});
however such proof does not imply termination.

\paragraph{Approximate specifications.}
  We point out usefulness of approximate specifications.
  They are crucial for avoiding unnecessary
  complications in constructing specifications and in
   correctness and completeness proofs.
  They are natural:  when starting construction of a program, 
  the relations it should compute are often known only approximately.
  Sections \ref{sec:logicProgram1}, \ref{sec:program} provide an example.
  Also, it is often cumbersome (and unnecessary) to exactly establish the
  relations computed by a program, 
  This is the case in our example, as discussed in Section
  \ref{sec:transformations}. 
  See \cite{drabent.tocl16} for further examples and discussion.

\paragraph{Program transformations.}
The program development example presented here differs from those
employing semantics-preserving transformations
\cite[and the references therein]{PettorossiPS10shorter}.
In our example,
  the semantics of (the common predicates in) the consecutive versions 
  of the program differ.  What is unchanged, is the correctness
  and completeness w.r.t.\ an approximate specification.
  In Section \ref{sec:transformations}
  we argue that the paradigm of semantics-preserving transformations
  is inapplicable to our case.
  This suggests introducing a more general paradigm of program transformations which 
  preserve correctness and completeness w.r.t.\ an approximate specification.

\paragraph{Final remarks.}
We are interested in declarative programming.
  The example presented here
  is intended to show how much of the programming task can be
  done declaratively, 
  without considering the operational semantics; how 
   ``logic'' could be separated from ``control.''  A substantial part of work
  could be done at the stage of a pure logic program.
  At this stage, correctness, completeness, termination, non-floundering, and
  occur-check freedom were formally proven;
  the reasoning about correctness and completeness was declarative.
  It is important that all the
  considerations and decisions about the program execution and efficiency
  (only superficially treated in this paper) are independent from those related
  to the declarative semantics,
  in particular to the correctness and completeness of the program.

  We argue that the employed proof methods 
  and the program construction approach 
  are simple, and
  correspond to a natural way of declarative thinking about programs.
  We believe that they can be actually used -- maybe at an informal level -- in
  practical programming; this is supported 
  (in addition to rather small examples in the previous work)
  by the main example of this article.
\paragraph{\bf Acknowledgement}
Extensive comments of anonymous referees were helpful in improving the
presentation.

\bibliographystyle{acmtrans}
\bibliography{bibshorter,bibpearl,bibmagic,bibcut}

\begin{thebibliography}{}

\bibitem[\protect\citeauthoryear{Apt}{Apt}{1997}]{Apt-Prolog}
{\sc Apt, K.~R.} 1997.
\newblock {\em From Logic Programming to {P}rolog}.
\newblock International Series in Computer Science. Prentice-Hall.

\bibitem[\protect\citeauthoryear{Apt and Luitjes}{Apt and
  Luitjes}{1995}]{AptL95.delays}
{\sc Apt, K.~R.} {\sc and} {\sc Luitjes, I.} 1995.
\newblock Verification of logic programs with delay declarations.
\newblock In {\em Algebraic Methodology and Software Technology, {AMAST} '95,
  Proceedings}, {V.~S. Alagar} {and} {M.~Nivat}, Eds. Lecture Notes in Computer
  Science, vol. 936. Springer, 66--90.

\bibitem[\protect\citeauthoryear{Apt and Pedreschi}{Apt and
  Pedreschi}{1993}]{AP93}
{\sc Apt, K.~R.} {\sc and} {\sc Pedreschi, D.} 1993.
\newblock Reasoning about termination of pure {P}rolog programs.
\newblock {\em Information and Computation\/}~{\em 106,\/}~1, 109--157.

\bibitem[\protect\citeauthoryear{Bezem}{Bezem}{1993}]{DBLP:journals/jlp/Bezem93}
{\sc Bezem, M.} 1993.
\newblock Strong termination of logic programs.
\newblock {\em J. Log. Program.\/}~{\em 15,\/}~1{\&}2, 79--97.

\bibitem[\protect\citeauthoryear{Carlsson and Mildner}{Carlsson and
  Mildner}{2012}]{sicstus.Carlsson.tplp12}
{\sc Carlsson, M.} {\sc and} {\sc Mildner, P.} 2012.
\newblock {SICStus Prolog} -- the first 25 years.
\newblock {\em {TPLP}\/}~{\em 12,\/}~1-2, 35--66.

\bibitem[\protect\citeauthoryear{Clark}{Clark}{1979}]{Clark79}
{\sc Clark, K.~L.} 1979.
\newblock Predicate logic as computational formalism.
\newblock Tech. Rep. 79/59, Imperial College, London. December.

\bibitem[\protect\citeauthoryear{Davis, Logemann, and Loveland}{Davis
  et~al\mbox{.}}{1962}]{DBLP:journals/cacm/DavisLL62}
{\sc Davis, M.}, {\sc Logemann, G.}, {\sc and} {\sc Loveland, D.~W.} 1962.
\newblock A machine program for theorem-proving.
\newblock {\em Commun. {ACM}\/}~{\em 5,\/}~7, 394--397.

\bibitem[\protect\citeauthoryear{Deransart}{Deransart}{1993}]{DBLP:journals/tcs/Deransart93}
{\sc Deransart, P.} 1993.
\newblock Proof methods of declarative properties of definite programs.
\newblock {\em Theor. Comput. Sci.\/}~{\em 118,\/}~2, 99--166.

\bibitem[\protect\citeauthoryear{Deransart and Ma{\l}uszy\'nski}{Deransart and
  Ma{\l}uszy\'nski}{1993}]{Deransart.Maluszynski93}
{\sc Deransart, P.} {\sc and} {\sc Ma{\l}uszy\'nski, J.} 1993.
\newblock {\em A Grammatical View of Logic Programming}.
\newblock The MIT Press.

\bibitem[\protect\citeauthoryear{Deville}{Deville}{1990}]{Deville}
{\sc Deville, Y.} 1990.
\newblock {\em Logic Programming: Systematic Program Development}.
\newblock Addison-Wesley.

\bibitem[\protect\citeauthoryear{Drabent}{Drabent}{2012}]{drabent12.iclp}
{\sc Drabent, W.} 2012.
\newblock Logic + control: An example.
\newblock In {\em Technical Communications of the 28th International Conference
  on Logic Programming (ICLP'12)}, {A.~Dovier} {and} {V.~S. Costa}, Eds.
  Leibniz International Proceedings in Informatics (LIPIcs), vol.~17. 301--311.
\newblock \small{\url{http://drops.dagstuhl.de/opus/volltexte/2012/3631}}.

\bibitem[\protect\citeauthoryear{Drabent}{Drabent}{2015}]{drabent.lopstr14}
{\sc Drabent, W.} 2015.
\newblock On completeness of logic programs.
\newblock In {\em Logic Based Program Synthesis and Transformation, LOPSTR
  2014. Revised Selected Papers}. Lecture Notes in Computer Science, vol. 8981.
  Springer.
\newblock Extended version in {\em CoRR} abs/1411.3015 (2014).
  \small{\url{http://arxiv.org/abs/1411.3015}}.

\bibitem[\protect\citeauthoryear{Drabent}{Drabent}{2016a}]{drabent.tocl16}
{\sc Drabent, W.} 2016a.
\newblock Correctness and completeness of logic programs.
\newblock {\em ACM Trans.\ Comput.\ Log.\/}~{\em 17,\/}~3, 18:1--18:32.

\bibitem[\protect\citeauthoryear{Drabent}{Drabent}{2016b}]{drabent.Herbrand.2016}
{\sc Drabent, W.} 2016b.
\newblock On definite program answers and least {Herbrand} models.
\newblock {\em {TPLP}\/}~{\em 16,\/}~4, 498--508.

\bibitem[\protect\citeauthoryear{Drabent}{Drabent}{2017}]{drabent.faoc16}
{\sc Drabent, W.} 2017.
\newblock Proving completeness of logic programs with the cut.
\newblock {\em Formal Aspects of Computing\/}~{\em 29,\/}~1, 155--172.

\bibitem[\protect\citeauthoryear{Drabent and Mi{\l}kowska}{Drabent and
  Mi{\l}kowska}{2005}]{DBLP:journals/tplp/DrabentM05shorter}
{\sc Drabent, W.} {\sc and} {\sc Mi{\l}kowska, M.} 2005.
\newblock Proving correctness and completeness of normal programs -- a
  declarative approach.
\newblock {\em TPLP\/}~{\em 5,\/}~6, 669--711.

\bibitem[\protect\citeauthoryear{Genaim and King}{Genaim and
  King}{2008}]{king.non-suspension2008}
{\sc Genaim, S.} {\sc and} {\sc King, A.} 2008.
\newblock Inferring non-suspension conditions for logic programs with dynamic
  scheduling.
\newblock {\em ACM Trans. Comput. Log.\/}~{\em 9,\/}~3.

\bibitem[\protect\citeauthoryear{Gomes, Kautz, Sabharwal, and Selman}{Gomes
  et~al\mbox{.}}{2008}]{handbook-SATsolvers}
{\sc Gomes, C.~P.}, {\sc Kautz, H.}, {\sc Sabharwal, A.}, {\sc and} {\sc
  Selman, B.} 2008.
\newblock Satisfiability solvers.
\newblock In {\em Handbook of Knowledge Representation}, {F.~van Harmelen},
  {V.~Lifschitz}, {and} {B.~Porter}, Eds. Chapter~2, 89--134.

\bibitem[\protect\citeauthoryear{Howe and King}{Howe and
  King}{2012}]{howe.king.tcs-shorter}
{\sc Howe, J.~M.} {\sc and} {\sc King, A.} 2012.
\newblock A pearl on {SAT} and {SMT} solving in {Prolog}.
\newblock {\em Theor. Comput. Sci.\/}~{\em 435}, 43--55.

\bibitem[\protect\citeauthoryear{King}{King}{2012}]{king.private2012}
{\sc King, A.} 2012.
\newblock Private communication.

\bibitem[\protect\citeauthoryear{Kowalski}{Kowalski}{1979}]{DBLP:journals/cacm/Kowalski79}
{\sc Kowalski, R.~A.} 1979.
\newblock Algorithm = logic + control.
\newblock {\em Commun. ACM\/}~{\em 22,\/}~7, 424--436.

\bibitem[\protect\citeauthoryear{Maher}{Maher}{1988}]{DBLP:books/mk/minker88/Maher88}
{\sc Maher, M.~J.} 1988.
\newblock Equivalences of logic programs.
\newblock In {\em Foundations of Deductive Databases and Logic Programming.},
  {J.~Minker}, Ed. Morgan Kaufmann, 627--658.

\bibitem[\protect\citeauthoryear{Nguyen, Schreye, Giesl, and
  Schneider{-}Kamp}{Nguyen
  et~al\mbox{.}}{2011}]{NguyenSGS11-termination-shorter}
{\sc Nguyen, M.~T.}, {\sc Schreye, D.~D.}, {\sc Giesl, J.}, {\sc and} {\sc
  Schneider{-}Kamp, P.} 2011.
\newblock Polytool: Polynomial interpretations as a basis for termination
  analysis of logic programs.
\newblock {\em {TPLP}\/}~{\em 11,\/}~1, 33--63.

\bibitem[\protect\citeauthoryear{Pedreschi and Ruggieri}{Pedreschi and
  Ruggieri}{1999}]{DBLP:journals/jlp/PedreschiR99}
{\sc Pedreschi, D.} {\sc and} {\sc Ruggieri, S.} 1999.
\newblock Verification of logic programs.
\newblock {\em J. Log. Program.\/}~{\em 39,\/}~1-3, 125--176.

\bibitem[\protect\citeauthoryear{Pedreschi, Ruggieri, and Smaus}{Pedreschi
  et~al\mbox{.}}{2002}]{PedreschiRS02.TPLP.terminating}
{\sc Pedreschi, D.}, {\sc Ruggieri, S.}, {\sc and} {\sc Smaus, J.-G.} 2002.
\newblock Classes of terminating logic programs.
\newblock {\em {TPLP}\/}~{\em 2,\/}~3, 369--418.

\bibitem[\protect\citeauthoryear{Pettorossi, Proietti, and Senni}{Pettorossi
  et~al\mbox{.}}{2010}]{PettorossiPS10shorter}
{\sc Pettorossi, A.}, {\sc Proietti, M.}, {\sc and} {\sc Senni, V.} 2010.
\newblock The transformational approach to program development.
\newblock In {\em A 25-Year Perspective on Logic Programming: Achievements of
  the Italian Association for Logic Programming, {GULP}}, {A.~Dovier} {and}
  {E.~Pontelli}, Eds. Lecture Notes in Computer Science, vol. 6125. Springer,
  112--135.

\bibitem[\protect\citeauthoryear{Pfenning}{Pfenning}{1992}]{types.lp.92-short}
{\sc Pfenning, F.}, Ed. 1992.
\newblock {\em Types in Logic Programming}.
\newblock The MIT Press.

\bibitem[\protect\citeauthoryear{Shapiro}{Shapiro}{1983}]{Shapiro.book}
{\sc Shapiro, E.} 1983.
\newblock {\em Algorithmic Program Debugging}.
\newblock The MIT Press.

\bibitem[\protect\citeauthoryear{Smaus, Hill, and King}{Smaus
  et~al\mbox{.}}{1998a}]{SmausHK98.LOPSTR98}
{\sc Smaus, J.}, {\sc Hill, P.~M.}, {\sc and} {\sc King, A.} 1998a.
\newblock Preventing instantiation errors and loops for logic programs with
  multiple modes using block declarations.
\newblock In {\em Logic Programming Synthesis and Transformation, LOPSTR'98,
  Proceedings}. 289--307.

\bibitem[\protect\citeauthoryear{Smaus, Hill, and King}{Smaus
  et~al\mbox{.}}{1998b}]{SmausHK98.PLILP98}
{\sc Smaus, J.}, {\sc Hill, P.~M.}, {\sc and} {\sc King, A.} 1998b.
\newblock Termination of logic programs with block declarations running in
  several modes.
\newblock In {\em Principles of Declarative Programming, PLILP'98,
  Proceedings}. 73--88.

\end{thebibliography}
\end{document}